\documentclass[fleqn,usenatbib]{mnras}

\usepackage{newtxtext,newtxmath}

\usepackage[T1]{fontenc}
\usepackage{pdflscape}	
\DeclareRobustCommand{\VAN}[3]{#2}
\let\VANthebibliography\thebibliography
\def\thebibliography{\DeclareRobustCommand{\VAN}[3]{##3}\VANthebibliography}

\usepackage{graphicx}	
\usepackage{amsmath}	

\title[Black Hole Mass Accretion Rates and Efficiency Factors]
{Black Hole Mass Accretion Rates and Efficiency Factors for over 750 AGN and Multiple GBH}

\author[R. A. Daly]{Ruth A. Daly\thanks{E-mail: rdaly@psu.edu}\\
Penn State University, Berks Campus, Reading, PA 19608, USA\\
Center for Computational Astrophysics, Flatiron Institute, 162 Fifth Avenue, New York, NY 10010, USA}

\date{Accepted 8 October 2020. Received 17 September 2020; in original form 24 July 2020}

\pubyear{2020}

\begin{document}
\label{firstpage}
\pagerange{\pageref{firstpage}--\pageref{lastpage}}
\maketitle

\date{Accepted XXX. Received YYY; in original form ZZZ}

\pubyear{2020}


\begin{abstract}
Mass accretion rates in dimensionless and physical units, and 
efficiency factors describing the total radiant luminosity of the disk and
the beam power of the outflow are 
studied here. Four samples of sources including 
576 LINERs, 100 classical double (FRII) radio sources, 80 relatively 
local AGN, and 103 measurements of four stellar mass X-ray 
binary systems, referred to as Galactic Black Holes (GBH),  
are included in the study. 
All of the sources have highly collimated outflows leading to 
compact radio emission or powerful extended (FRII) radio emission.
The properties of each of the full samples are explored, as are 
those of the four individual GBH, and sub-types of the FRII and local AGN samples. 
Source types and sub-types that 
have high, medium, and low values of accretion rates and efficiency factors 
are identified and studied.  
A new efficiency factor that describes the relative impact of black hole spin  
and mass accretion rate on the beam power is defined and studied, 
and is found to provide a new and interesting diagnostic. 
Mass accretion rates for 13 sources and efficiency factors for 6 sources 
are compared with values obtained independently,   
and indicate that similar values are obtained with independent 
methods. The mass accretion rates and efficiency factors obtained here 
substantially increase the number of values available, and improve our 
understanding of their relationship 
to source types. The redshift dependence of quantities is presented 
and the impact on the results is discussed. 
\end{abstract}




\section{Introduction} \label{sec:intro}
Mass accretion rates and associated efficiency factors are 
important quantities that describes the state of 
stellar and supermassive black hole systems. They  
provide information regarding  
processes that lead to the accretion, the state of the accretion disk, 
and how outflows and winds from accreting systems can affect their
environment. Thus, they are important diagnostics of the state of the black 
hole system (e.g. Heckman \& Best 2014; Yuan \& Narayan 2014). 
Accretion rates and efficiency factors have been determined using a 
variety of methods and diagnostics, and this has allowed estimates of 
these quantities for sources with a broad range of properties  
(e.g. Bian \& Zhao 2003; Davis \& Laor 2011; Raimundo et al. 2012; 
Wu, Lu, \& Zhang 2013; Trakhtenbrot 2014; Netzer \& Trakhtenbrot 2014; 
Trakhtenbrot, Volonteri, \& Natarajan 2017; Jiang et al. 2019; O'Dea \& 
Saikia 2020).

Here, the new method of obtaining mass accretion rates introduced by Daly (2019) 
is applied to obtain and study accretion rates and efficiency factors. 
The "outflow method," developed and applied  by Daly (2016, 2019) 
(hereafter D16 and D19) and 
Daly et al. (2018), (hereafter D18) applies to black hole systems that include a 
black hole, accretion disk, and a highly collimated outflow. 
The highly collimated outflow must lead to radio emission from a compact radio 
source or from a powerful extended radio source that can be used to 
study the properties of the highly collimated outflow, as described in section 2.
The method 
allows empirical determinations of accretion disk 
properties such as the mass accretion rate and disk magnetic field strength
in dimensionless and 
physical units.

\subsection{Overview of the Outflow Method}

The outflow method of determining and studying black hole spin and 
accretion disk properties was proposed and applied by D16 and 
D19, and followed the methods developed and applied by Daly (2009, 2011) 
and Daly \& Sprinkle (2014) 
to estimate black hole spin values based on outflow beam powers. 
A detailed overview of the method, it's development, 
and it's application is provided in section 1.1 of D19.  The method is 
empirically based, and includes one key theoretical input, 
as described below. To understand the method and it's 
application it is important to relinquish preconceived notions 
and assumptions that are based on the application 
of specific accretion disk or other  
models and focus on what is indicated empirically. 

As is often the case with 
theoretical developments, with hindsight, simple arguments that are consistent with the original method become 
evident, even though this is not the way the method was developed or the results obtained. 
In this spirit, a heuristic path to understand the method and its application is described here.  
This provides an alternative path to arrive at the same conclusions even though 
this is not the way the equations were derived, which is explained in detail in section 1.1. of D19. 

Consider a black hole
with irreducible mass $M$ and dimensionless spin parameter $j$ that has 
an accretion disk accreting 
matter at a rate $(dM/dt)$ in solar masses per year. A priori, these 
three quantities are independent. Given that the accretion disk of the 
system produces some bolometric luminosity, $L_{bol}$, the radiant efficiency 
of the disk may be determined empirically when $\dot{M} \equiv (dM/dt)$ is known. 

The outflow method has one key theoretical input, and the introduction of this 
theoretical input is motivated empirically. It is motivated by the fact that 
there are numerous black hole systems with outflow beam powers that exceed the 
bolometric luminosity of the disk by one to two orders of magnitude. 
For example, the results of D18 indicate that there are numerous black hole 
systems with beam powers that are 10 - 100 times larger than the bolometric 
luminosity of the accretion disk of the system (see Fig. 2 of D18). In these 
cases, it is reasonable to suppose that the highly collimated outflow is 
powered at least in part by the black hole spin   
(see also Narayan \& McClintock 2012, and Daly 2011). 
D18 also showed that the slopes and normalizations of the relationship between 
$\rm{Log} (L_j/L_{bol})$ and $\rm{Log}(L_{bol}/L_{Edd})$ 
are consistent within uncertainties for the four samples studied, even 
though these quantities have a broad range of values of 
about $(10^{-2}$ to $10^2)$ for $(L_j/L_{bol})$ and  
$(10^{-6}$ to $1)$ for $(L_{bol}/L_{Edd})$, where $L_j$ is the beam power and $L_{Edd}$ is the 
Eddington luminosity. This suggests that the sources in the four samples studied 
by Daly et al. (2018) are 
governed by the same physics, and therefore, all of the outflows are 
powered at least in part by black hole spin. 

This indicates that the functional form of the 
theoretical equation that describes spin powered outflows 
(e.g. Blandford \& Znajek 1977; Blandford 1990; Meier 1999) 
may be compared with the functional form of empirically determined relationships 
(D16, D18, D19). 
Once the theoretical equation is written in dimensionless separable form 
(see eq. 1), terms in the empirically determined relationship may be 
identified with terms in the theoretical equation to solve for 
the properties of the accretion disk and black hole spin. 
Eq. (1) is the primary theoretical input to the outflow method. 
Other aspects of the method are empirically based. No particular accretion disk 
emission model is assumed, and no detailed jet power/emission 
model is adopted. 

D19 showed that the fundamental equation that 
describes outflows that are powered at least in part by black hole spin,
$L_j \propto B_p^2 M_{dyn}^2 f(j)$ (e.g. Blandford \& Znajek 1977; 
Blandford 1990; Meier 1999) is separable and may be written as 
\begin{equation}
(L_j/L_{Edd}) = g_j ~(B/B_{Edd})^2 ~F^2
\end{equation}
(see eq. 6 from D19). Here $L_j$ is the beam power of the outflow, or  
the energy per unit time carried away from the black hole region in the 
form of a collimated outflow, typically thought to be in the form 
of directed kinetic energy, $B_p$ is the poloidal component of the 
accretion disk magnetic field, 
$B$ is the disk magnetic field strength, $B_{Edd}$ is the Eddington magnetic 
field strength (e.g. Rees 1984; Blandford 1990; Dermer, Finke, \& Menon 2008), 
$B_{Edd} \approx 6 (M_{dyn}/10^8 M_{\odot})^{-1/2} \times 10^4 ~\rm{G}$,  
$M_{dyn}$ is the total black hole mass, the normalized spin function 
is $F^2 \equiv (f(j)/f_{max})$ where $f(j)$ is the spin function and $f_{max}$ 
is the maximum value of this function (see D19),  
$g_j$ is the normalization factor for the beam power $L_j$ in units of the 
Eddington Luminosity, $L_{Edd}$, $L_j/L_{Edd}(max) = g_j$. 
It is clear that this is separable: information regarding the outflow is described by 
$L_j/L_{Edd}$ and $g_j$; that related to black hole spin is described 
by $F^2$; and, as explained below, that related the accretion event is described by the term 
$(B/B_{Edd})^2 = dm/dt=(L_{bol}/(g_{bol} L_{Edd}))^A$ (see eq. 7 of D19), 
where the value of $A$ is obtained for each sample 
from Table 2 of D18, as summarized by D19, and is typically $A \simeq 0.45$. 
Note that the ratio $(B_p/B)^2$ is absorbed into the normalization factor $g_j$. 
The method was further developed and applied by D19 to study black hole 
spin functions and spin, the magnetic field strength of the accretion disk, 
$B$ in physical units, and 
the magnetic field strength or pressure of the accretion disk in the 
dimensionless 
form $(B/B_{Edd})$. This work did not explicitly compute mass 
accretion rates in physical units 
or efficiency factors. These topics are addressed in this paper. 

The identification of terms in eq. (1) with 
empirically determined quantities was carried out with detailed calculations, 
presented and discussed by D16, D18, and D19. 
With hindsight a simple identification of terms 
becomes evident. The empirically determined relationship between beam power and 
disk luminosity, which by construction 
does not include a term related to black hole spin, 
has the form $(L_j/L_{Edd}) \propto (L_{bol}/L_{Edd})^A,$
where the only term related to the overall properties of the 
accretion disk is $(L_{bol}/L_{Edd})^A$. In the theoretical 
equation given by eq. (1) there is only one term that describes 
the overall properties of the 
accretion disk, and that is the term $(B/B_{Edd})^2$. Thus,
identifying these terms, and noting that each is normalized to have a 
maximum value of unity, it is easy to see that 
\begin{equation}
(B/B_{Edd})^2 = (L_{bol}/L_{Edd})^A~, 
\end{equation}
as originally indicated by the detailed calculations discussed by D19.  

To see how this is related to mass accretion rate, consider 
standard representations (e.g. Yuan \& Narajan 2014; Ho 2009) 
for the bolometric disk luminosity
\begin{equation}
L_{bol} \propto \epsilon_{bol} (dM/dt) \propto \epsilon_{bol}~ \dot{m} ~L_{Edd},
\end{equation}
and the beam power
\begin{equation}
L_j \propto (dM/dt)~ F^2 \propto  \dot{m} ~L_{Edd} ~F^2 
\end{equation}
where the efficiency factor 
$\epsilon_{bol}$ that describes the radiant luminosity of the disk 
is not assumed to be constant, the dimensionless mass accretion rate 
is defined as 
$\dot{m} \equiv \dot{M}/\dot{M}_{Edd}$, 
and the Eddington mass accretion rate is defined as 
$\dot{M}_{Edd} \equiv L_{Edd}/c^2$. 
(Some authors adopt different definitions of the Eddington mass accretion rate 
and the dimensionless mass accretion rate. It is shown in 
section 3.3 that consistent results are obtained with 
those definitions and the definitions adopted here.) 
Though eq. (4) is most frequently adopted for sources with low values of 
$(L_{bol}/L_{Edd})$, it is reasonable to suppose that the same relationship holds 
for all of the sources since the samples studied include many sources 
with low values of $(L_{bol}/L_{Edd})$, and all of the samples have 
consistent values of the slope and normalization relating  $(L_j/L_{Edd})$  
to $(L_{bol}/L_{Edd})$ (e.g. D18). Combining eqs. (1), (2), (3), and (4), we obtain 
\begin{equation}
    \epsilon_{bol} = (dm/dt)^{(1-A)/A}   
\end{equation}
where the constant of proportionality must be unity because  
$\epsilon_{bol}$ and $(dm/dt)$ are each normalized to have a maximum 
value of one. 
Thus, $\epsilon_{bol} \simeq (dm/dt)^{1.2}$ or $\epsilon_{bol} \simeq (dm/dt)$
for $A \simeq 0.45$ and $A \simeq 0.5$, respectively (see also Ho 2009). 
Note that in D16, a more general functional form for $L_j$ was adopted, 
and the same results were obtained.  
For the types of sources studied here, 
$\epsilon_{bol}$ is only a constant when $A = 1$, which is quite clearly ruled 
out for these sources (D16, D18, D19). Substituting 
eq. (3) into eq. (2), and applying eq. (5), we obtain
\begin{equation}
    (B/B_{Edd})^2 = (\epsilon_{bol} \dot{m})^A~ = ~\dot{m}~.  
\end{equation}
Thus,
\begin{equation}
    dM/dt = \dot{m}~ L_{Edd}~c^{-2} = (L_{bol}/L_{Edd})^A~L_{Edd}~c^{-2}
\end{equation}
so $dM/dt$ can be empirically determined using eq. (7). 

It is interesting to note that empirical results indicate $(dm/dt)$ and thus 
$\epsilon_{bol}$ are independent of the black hole spin function $F$ and 
black hole spin (D16, D19). 
This is also suggested by the fact that there is no 
empirical indication that the relationship between $(L_j/L_{bol})$ and 
$(L_{bol}/L_{Edd})$ changes as $(L_{bol}/L_{Edd}) \rightarrow 1$, 
for the sources studied, as discussed 
by D16 and D19. 

It is interesting to note that eq. (2) provides an estimate of the 
gas pressure of the accretion disk in dimensionless form (see section 3.3. 
of D19). And the magnetic field strength of the disk, which is  
obtained from eq. (2), provides an estimate of the disk pressure 
in physical units (see section 3.1 of D19).

Here, four samples of sources, including 756 AGN and 103 measurements 
of four stellar-mass black holes associated with X-ray binaries are considered. 
The samples, source parameters, and parameter uncertainties 
are described in section 2. 
Methods of obtaining mass accretion rates and efficiency factors are 
described in sections 3.1 and 3.2, respectively. The results are 
discussed in section 4.1 and 4.2, respectively. 
An overview of the different source parameters associated with the accretion 
disk that can be obtained with the method, and the impact of the redshift range of 
the sources on the results are discussed in section 5. The conclusions are 
summarized in section 6. 

\section{Data} \label{sec:data}

Four samples are considered here. The samples include 576 LINERs 
from Nisbet \& Best (2016) (hereafter NB16); 100 classical double (FRII) 
sources from D16 and D19; 80 local AGN that are compact radio sources from 
Merloni et al. (2003) (hereafter M03); and 103 observations of four stellar-mass Galactic 
Black Hole systems (GBH) that are in X-ray binary systems. The stellar-mass GBH data  
includes the 102 observations listed in Table 2 of D19 plus 
one observation of A0 6200 discussed in section 5 of D19 (the data for the second 
observation of A0 6200 is from Gou et al. 2010; Kuulkers et al. 1999; and  
Owen et al. 1976). These are stellar mass X-ray binaries, and each of the GBH 
have multiple simultaneous radio and X-ray observations (Saikia et al. 2015) 
(hereafter S15) who obtained 
the data from Corbel et al. (2013), Corbel, K\"{o}rding, \& Kaaret (2008),  
Zycki, Done \& Smith (1999), Shahbaz et al. (1996), Merloni et al. (2003), 
Gallo et al. (2006), Gelino, Harrison \& Orosz (2001), 
Shahbaz, Naylor \& Charles (1994), and Jonker \& Nelemans (2004). 
These four samples were studied by D19 who reported spin 
functions, spin values, and accretion disk magnetic field strengths in 
dimensionless and physical units for each source and observation. 

The samples were selected for the purpose of studying black hole spin
using source outflow properties following the work of Daly (2009, 2011), 
Narayan \& McClintock (2012), and Daly \& Sprinkle (2014), for example. 
It has long been thought that powerful 
collimated outflows from black hole systems are likely to be 
powered by black hole spin (e.g. Blandford \& Znajek 1977; 
Rees 1984; Begelman, Blandford, \& Rees 1984; Blandford 1990). 

Individual samples, each comprised of a specific type of source  
including the NB16 LINERs, D16/D19 FRII (classical double) sources, 
M03 local AGN sample, and the S15 stellar-mass GBH were selected. The four different 
types of black hole systems allow studies of and 
comparisons between four different categories of black hole system, 
including supermassive and stellar-mass systems. 
Another criteria was that reliable determinations of the 
bolometric disk luminosity, beam power, and black hole mass were 
available. 

The NB16 LINERS, M03 AGN, and S15 GBH all lie on the fundamental 
plane of black hole activity. Thus, black hole masses were available for 
all of the sources, bolometric disk luminosities were obtained from 
the (2 - 10) keV X-ray luminosity using a standard conversion 
(e.g. Ho 2009; D18), and the beam powers were obtained by rotating the 
fundamental plane to the fundamental line of black hole activity 
using eq. (4) of D18 and the values of C and D listed in Table 1 of 
D18 for each sample. 
Note that time variability of the mass accretion rate and it's impact 
on various source properties is discussed in detail in sections 
5 and 6 of D19. 
For the FRII sources, the black hole masses were obtained from 
McLure et al. (2004, 2006). The bolometric disk 
luminosities were obtained
from the O[III] luminosities of Grimes, Rawlings \& Willott (2004) 
using a standard conversion 
(e.g. Heckman et al. 2004; Dicken et al. 2014). The beam powers 
were obtained using the strong shock method (described in detail in 
section 1.1 of D19) for 12 of the sources (O'Dea et al. 2009), and 
the relationship between beam power and 178 MHz radio luminosity obtained  
by Daly et al. (2012) using 31 sources with beam powers determined 
with the strong shock method (O'Dea et al. 2009). The source types are from Grimes, Rawlings \& Willott (2004); W sources have weak emission lines, and Q sources are quasars.

Uncertainties of all quantities used to determine  
the black hole mass, bolometric luminosity, and beam power were included to  
obtain the uncertainty of each of these quantities, as described in section 2 of 
D19. The uncertainties $\delta \rm{Log}(L_{Edd})$, 
$\delta \rm{Log}(L_{bol})$, $\delta \rm{Log}(L_j)$ listed in section 2 of D19 
are propagated to obtain uncertainties of all the 
quantities discussed in this paper using a method identical to that described 
in detail in section 3.3 of D19. For the GBH, the uncertainty of the black hole 
mass is small compared with uncertainties of the bolometric disk luminosity and 
the beam power, so uncertainties of the black hole mass 
are neglected for the GBH. In addition, the black hole mass of a given GBH is 
constant during the outflow events, and the black hole mass range for the GBH is 
small, so neither the dispersion of the GBH black hole mass nor the uncertainty of GBH  
black hole masses are included in Table 1. 

Here, the dimensionless disk magnetic field strength, that is, the 
ratio of the accretion disk field strength to the Eddington 
magnetic field strength, is used to obtain the mass accretion rate 
in dimensionless and physical units for each system and for each observation  
of the GBH. The mass accretion rate of each source is combined with the 
bolometric disk luminosity, outflow beam power, and black hole spin 
function to define and study three efficiency factors associated with each 
black hole system, and the properties of different samples and types of 
sources are summarized in Table 1. 
Mass accretion rates and bolometric efficiency factors 
obtained here are compared with those obtained independently 
in Table 2. Results for individual sources and observations are listed in 
Tables 3 - 6, and provide a measure of the state of the system at the time 
of the observation, since any given source may be time variable. 
Quantities are obtained in the context of a spatially 
flat cosmological model 
with mean mass density of non-relativistic matter relative to the critical 
density today of $\Omega_m = 0.3$, a similarly normalized cosmological 
constant of $\Omega_{\Lambda} = 0.7$, and a value of Hubble's constant of 
70 km/s/Mpc. 

\section{Method} \label{sec:method}
\subsection{The Mass Accretion Rate in Dimensionless and Physical Units}

The mass accretion rate of each source in 
dimensionless and physical units can be empirically determined using the 
results obtained by D19 who showed that the dimensionless mass accretion 
rate $(dm/dt) = (B/B_{Edd})^2$, where B is the strength of the magnetic field 
anchored in the accretion disk and $B_{Edd} \propto M_{dyn}^{-1/2}$ is the 
Eddington magnetic field 
strength (e.g. Rees 1984); D19 list empirically determined values 
of $(B/B_{Edd})^2$ and $B$ for each source. Given the definition of the dimensionless mass 
accretion rate, $(dm/dt) \equiv (dM/dt) c^2/L_{Edd}$, it is easy to see that
\begin{equation}
    \left({dM \over dt}\right) = \left({B \over B_{Edd}}\right)^2 ~\left({L_{Edd}\over c^2}\right)~,
\end{equation}
where $L_{Edd}$ is the Eddington luminosity of the hole, $L_{Edd} \simeq 1.3 \times 10^{38} (M_{dyn}/M_{\odot}) \hbox{ erg s}^{-1}$, 
so the mass accretion rate in physical units $dM/dt$ can be obtained from $dm/dt$. 
The results are summarized in Table 1. 

A comparison between mass accretion rates obtained here 
with those obtained independently are discussed in section 4.1 and 
are shown in Table 2. 

Table 1 includes the mean value and dispersion of each quantity, with the 
dispersion measured by the one sigma standard deviation of the sample or sub-sample. 
The uncertainty per source of each quantity is included in (brackets) following 
the mean value and dispersion of each quantity in the top part 
of Table 1. The uncertainty per source of each quantity 
is obtained by propagating the uncertainties  
$\delta \rm{Log}(L_{bol})$, $\delta \rm{Log}(L_j)$, $\delta \rm{Log}(L_{Edd})$ 
of the input parameters adopting a value of $A \approx 0.45$ for all samples, 
as described in section 2. 

\subsection{Accretion Disk and Outflow Efficiency Factors}

The efficiency factor of the accretion disk, $\epsilon_{bol}$, is defined as 
\begin{equation}
    \epsilon_{bol} \equiv {L_{bol} \over (c^2 ~dM/dt)}~. 
\end{equation}
For the sources studied here, $L_{bol}$ is determined from the 
[OIII] for the FRII sources, and from the (2-10) keV X-ray luminosity for the 
other samples (see D16 and D18), and the values of $L_{bol}$ listed in Tables 2-5 from 
D19 are applied here. 
The mass accretion rate $dM/dt$ is obtained as described in sections 1.1 and 3.1. 

A comparison between bolometric efficiency factor values, $\epsilon_{bol}$,
obtained here with those obtained independently are discussed in section 4.2 and are 
shown in Table 2.

The efficiency factor of the outflow, $\epsilon_j$, is defined as 
\begin{equation}
    \epsilon_j \equiv {L_{j} \over (c^2 ~dM/dt)} ~= g_j~ F^2.
\end{equation}
Outflow beam powers are based on the strong shock method for the 
FRII sources and on the mapping from the fundamental plane of black hole 
activity to the fundamental line of black hole activity for the three other 
samples, as described by D18 and D19 (see also O'Dea et al. 2009). 
The values of $L_j$ listed in Tables 2-5 from D19 are applied here. 

A new efficiency factor that measures the relative impact 
of black hole spin and the accretion disk on the Eddington-normalized 
beam power, $\epsilon_{s/d}$, is referred to as the  "spin relative to disk" efficiency factor. This efficiency factor is 
\begin{equation}
    \epsilon_{s/d} \equiv {F^2 \over (dm/dt)}.
\end{equation}
This ratio is interesting 
because beam power in Eddington units may be written as, 
$L_j/L_{Edd} = g_j (dm/dt) F^2$, 
so the ratio $F^2/(dm/dt)$ measures the relative impact of 
black hole spin and the dimensionless mass accretion rate on the Eddington 
normalized outflow beam power (see eq. 4).

The sample or sub-sample mean value and dispersion is listed in 
Table 1 for each quantity, and the uncertainty per source is included in brackets 
in top part of Table 1, as described in section 2 and 3.1. 

\subsection{Comparison with other Representations of Key Equations}

Throughout these investigations, 
the total radiant luminosity of the accretion disk is represented in the standard form
$L_{bol} \propto \epsilon_{bol} \dot{M}$,  
where the efficiency factor 
$\epsilon_{bol}$ is not assumed to be constant (see section 1.1). 
The dimensionless mass 
accretion rate and Eddington mass accretion rate are
defined to be 
$\dot{m} \equiv \dot{M}/\dot{M}_{Edd}$ and  
$\dot{M}_{Edd} \equiv L_{Edd}/c^2$, respectively. 

An equivalent representation can be obtained by absorbing the 
efficiency factor into the definition of the Eddington mass accretion rate 
(e.g. Yuan \& Narayan 2014). 
If the Eddington mass accretion rate is defined as 
$\dot{M}^{\prime}_{Edd} \equiv L_{Edd}/(\epsilon_{bol} c^2)$, 
and the dimensionless mass accretion rate is defined as 
$\dot{m}^{\prime} \equiv \dot{M}/\dot{M}^{\prime}_{Edd}$, then 
$L_{bol}$ becomes 
$L_{bol} \propto \dot{m}^{\prime} M_{dyn}$, with  
$\dot{m}^{\prime} = \epsilon_{bol} \dot{m}$. 

Since both D16 and D19 empirically determined that the 
efficiency factor $\epsilon_{bol}$ can not be constant for the 
sources studied, the first representation given by eq. (3), which 
allows explicit tracking of this efficiency factor, was selected.

\section{Results} 
\label{sec:results} 
Mass accretion rates in dimensionless and physical units obtained 
as described in section 3.1 and efficiency factors obtained as described 
in section 3.2 are summarized in Table 1. The mean value and  
dispersion (i.e. the one sigma standard deviation)  
for the sample, sub-sample, or source listed in column (1) are provided for each quantity. 
The estimated uncertainty per source of each quantity is indicated in brackets 
following the mean value and dispersion in the top part of the table, 
obtained as described in section 2 and 3.1. The input values used to compute 
each quantity are obtained from D19; for completeness, black hole mass is included 
in the Tables presented here. 

A comparison of mass accretion rates and bolometric efficiency factors obtained 
independently is provided in Table 2. 

Mass accretion rates, efficiency factors, and black hole masses 
for each source and measurement for the S15 sample of GBH, the D16/D19 sample of 
classical double (FRII) sources, the M03 sample of local AGN, and the 
NB16 sample of LINERs are listed in Tables 3, 4, 5, and 6, respectively. 

\subsection{Mass Accretion Rates}

\begin{figure}
    \centering
    \includegraphics[width=\columnwidth]{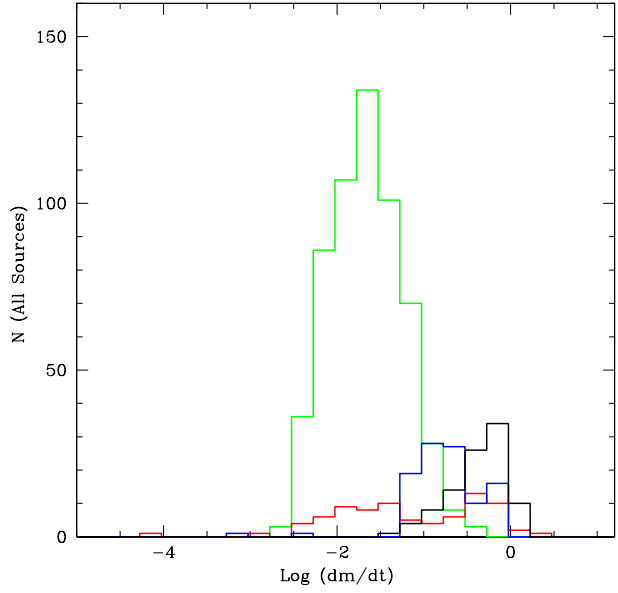}
\caption{Histograms of the dimensionless mass accretion 
rate, $\dot{m} \equiv dm/dt$. 
Here and throughout the paper, the NB 16 sample of 576 LINERs 
is shown in green; the D16/D19 sample of 100 FRII sources 
is shown in black; the M03 local sample of 80 AGN is shown 
in red; and the S15 sample of 103 measurements of 4 GBH is shown in blue.}
		  \label{fig:F1A}
    \end{figure}

\begin{figure}
    \centering
    \includegraphics[width=\columnwidth]{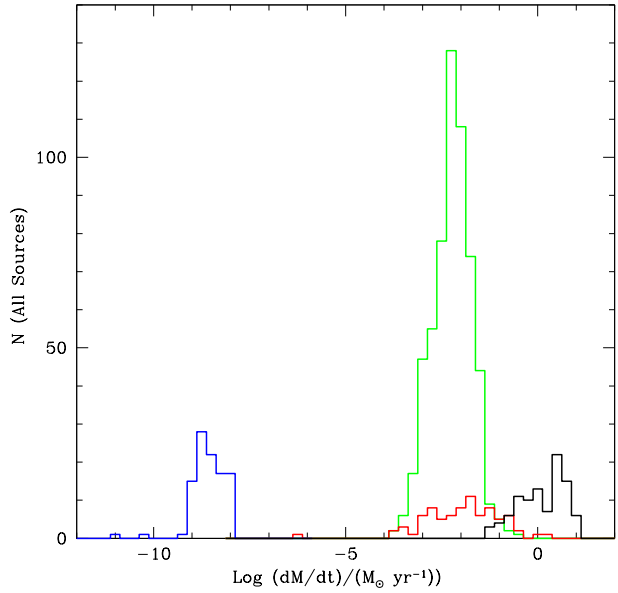}
\caption{Histograms of the mass accretion rate, 
$dM/dt$, in physical units of solar masses per year; sample colors are as in Fig. 
\ref{fig:F1A}.}
		  \label{fig:F3A}
    \end{figure}

Histograms of mass accretion rates in dimensionless and physical units 
are shown in Figs. \ref{fig:F1A} - \ref{fig:F3A}. 
The mean value and one sigma dispersion of each sample is listed in 
the first part of Table 1. The sample is indicated in column (1), 
the source type is indicated in column (2), the number of sources 
(or number of measurements for GBH)  
included in the mean and dispersion is listed in column (3), mean 
values for the mass accretion rate in dimensionless and physical units
are listed in columns (4) and (5), respectively. 

\begin{figure}
    \centering
    \includegraphics[width=\columnwidth]{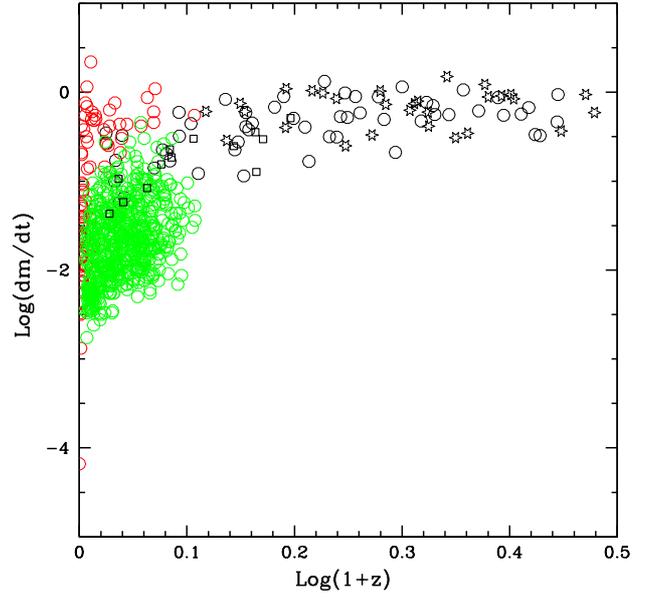}
\caption{The dimensionless mass accretion rate, $dm/dt$, is shown as a function 
of $(1+z)$ for the AGN samples. Classical double 
(FRII) LEG, HEG, and quasar (Q) source types are shown in black 
as open squares, open circles, and open stars, respectively. M03 and NB16 sources 
are shown as red and green open circles, respectively. }
		  \label{fig:F2A}
    \end{figure}

\begin{figure}
    \centering
    \includegraphics[width=\columnwidth]{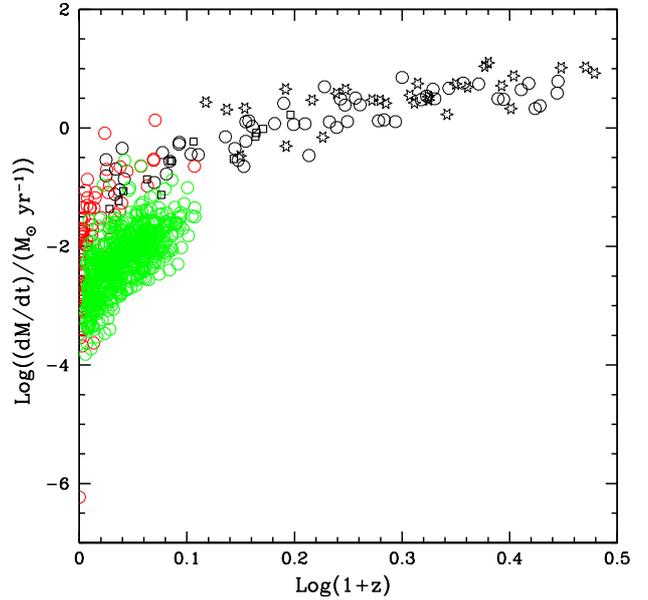}
\caption{The mass accretion rate, $dM/dt$, in physical units of 
solar masses per year is shown as a function of $(1+z)$ for the AGN samples; 
colors and symbols are as in Fig. 
\ref{fig:F2A}.}
		  \label{fig:F4A}
    \end{figure}

For AGN the dimensionless mass accretion rates are related to AGN type, 
as is evident from the values listed in the first four lines of Table 1, 
as expected based on the study by Ho (2009). The 100 FRII sources have 
the highest mass accretion rates for rates expressed in both dimensionless 
and physical units. Part of the reason for this is that the sample includes 
sources with redshifts between 
about zero and two, as illustrated in Figs. \ref{fig:F2A} - \ref{fig:F4A}, and 
the mass accretion rate in physical units increases with redshift. 
The M03 sample of local AGN has the second highest mass accretion rate, though 
there is substantial overlap between the M03 and NB16 samples for the mass 
accretion rate in physical units, and between the M03 sample and both the 
D19 and NB16 samples for the rate in dimensionless units. 
The mean dimensionless mass accretion rate for the GBH is similar to that 
for the AGN and, of course, the rate in physical units is much smaller than that 
obtained for AGN. The maximum values of the dimensionless mass accretion rate are 
close to unity, indicating that all of the sources studied here 
are accreting at Eddington or sub-Eddington levels. 
This limit is not externally imposed. 

It is clear from Figs. \ref{fig:F2A} - \ref{fig:F4A} that sources with lower
values of $dm/dt$ and $dM/dt$, which are evident at low redshift, 
drop out of each sample as redshift increases. 
This effect is particularly striking 
for $dM/dt$ and is likely a selection effect whereby sources with 
lower bolometric disk luminosities or lower radio luminosities 
that are at larger distances and redshift 
require observations to lower flux densities to be detected, and thus in 
flux limited surveys these sources   
drop out of the sample as redshift increases. 
Thus, the different distributions of $dM/dt$ seen in Fig. \ref{fig:F4A} is 
affected by these redshift selection effects. However, the upper envelope of 
the distributions seen in Fig. \ref{fig:F4A} indicate the maximum mass 
accretion rates for each of the source types as a function of redshift.

Some of the sources studied here have mass accretion rates obtained using 
independent methods, such as those discussed by Jiang et al. (2019) 
and Raimundo et al. (2012). These studies are based on a detailed 
accretion disk model of each source. The model includes numerous accretion disk 
parameters in addition to the mass accretion rate. 
In addition, the bolometric accretion disk 
luminosity is based on the optical properties of the accretion disk, and a 
detailed model of the disk, as described below. Thus, 
the method and it's application are quite different from those applied here, 
which are summarized in sections 1.1, 2, 3.1, and 3.2.
There are 14 independently determined values of 
mass accretion rates that overlap with 13 sources studied here 
(one source is included twice), and these are listed in Table 2. In Table 2, 
J19 refers to Jiang et al. (2019), (T5) refers to Table 5 (presented here and included below), 
R12(T1), R12(T2) and R12(T3) refer to Tables 1, 2 and 3 of Raimundo et al. 
(2012), respectively; the comparison of efficiency factors is 
discussed in section 4.2.  
The methods used by Jiang et al. (2019) and Raimundo et al. (2012) are based on 
optical observations and have completely different functional forms compared 
with those applied here (see eq. 3 of Raimundo et al. 2012; see also Davis \& 
Laor 2011; and Netzer \& Trakhtenbrot 2014), so their mass accretion rate and 
efficiency factor determinations are independent of those obtained here.

Values of the mass accretion rate obtained with independent methods are similar, 
within factors of few, for 12 of the 14 comparisons (see Table 2); 
two sources differ by about a factor of ten. 
Given the very different way that the accretion rates are obtained, 
this is agreement suggests that both methods provide reliable estimates
of mass accretion rates. Note that time variability may be a factor, as 
observations used to determine the accretion rate of a given source with 
different methods were obtained at different times. Thus, it is likely that  
each method provides mass accretion rate estimates that are accurate to factors 
of about a few. 

The second part of Table 1 lists mean values and dispersions obtained with 
multiple simultaneous radio and X-ray measurements of each GBH. 
For the stellar-mass GBH, the dimensionless mass accretion rates range are about 
$0.2$ for GX-339-4; $0.1$ for V404 Cyg; $0.08$ for J1118+480; and $0.01$ for 
A0 6200. These values are consistent with expectations for black hole 
X-ray binaries in the low-hard state 
(e.g. Skipper \& McHardy 2016), and are similar to the 
accretion and mass transfer rates from the companion star to the black 
hole discussed by McClintock et al. (2003) and Shao \& Li (2020). 
The mean accretion rate in physical units 
for 103 measurements of 4 stellar-mass GBH is about 
$3 \times 10^{-9} M_{\odot} yr^{-1}$, with three of the 4 stellar-mass GBH having 
quite similar values, and one source, A0 6200, having a lower mean value  
that results from two very different accretion rates of about 
$5 \times 10^{-9} M_{\odot} yr^{-1}$ and 
$10^{-11} M_{\odot} yr^{-1}$ (see Table 3). 
The input values for the GBH were adopted from S15, who 
considered GBH with simultaneous radio and X-ray data that were in 
the low-hard state (see S15 for details), thus the mass 
accretion rates obtained here apply to rates during that phase. 
The typical values obtained here based on 76 simultaneous 
radio and X-ray measurements of the source for GX 339-4 fall into a range that is 
intermediate between previously reported values 
(e.g. Zdziarski, Ziolkowski, \& Mikolajewska, 2019),  
which are obtained with methods independent of those introduced by D16 and D19, 
and applied here. Similarly, values obtained 
here for A0 6200 and V404 Cyg are within the range of values for these sources 
discussed by Lasota (2000) and Ziolkowski \& Zdziarski (2018). 
In all cases, the values obtained here 
only depend upon the identification of the term $(B/B_{edd})^2$ from eq. (1) 
with the empirically determined quantity $(L_{bol}/L_{Edd})^A$ 
(see eq. 2) as 
discussed in detail by D19, and thus are independent of a detailed accretion 
disk or mass transfer model for the systems, so the agreement is encouraging. 

Mass accretion rates for sub-types of AGN for the FRII sources and the M03 
sources are included in the third and fourth sections of Table 1;  
the sources from the NB16 sample are all one type of AGN, LINERs, 
so results from sub-types of the FRII and M03 samples can 
be compared with 
those for the LINERS. For the FRII sources, the quasars (Q)
and W galaxies, which have weak emission lines
as defined by Grimes et al. (2004), 
have the highest dimensionless mass accretion rates, though 
within the one sigma standard deviation for the sub-sample 
the rates are similar for the different types of
sources. For the FRII sources, the rates in physical units of $M_{\odot} yr^{-1}$, 
quasars have the highest accretion rates, followed by HEG, W, and 
LEG sources (e.g. Mingo et al. 2014; Fernandes et al. 2015). 

For the M03 sample, the 
quasars and NS1 sources have the highest dimensionless mass accretion rates, 
followed by S(1-1.9) and S2 sources, and L1.9 - L2 sources (with the source
types defined by M03). Mean values and the dispersions of accretion rates in 
both dimensionless and physical units for the M03 L type sources 
are similar to that obtained for the 576 LINERs in the NB16 sample. 
Mass accretion rates in physical units of all of the M03 source types are similar 
to each other and to that of the NB16 LINER sample, with the exception of the 
M03 quasars, which have rates about an order of magnitude larger than the other 
M03 source types studied here. 

\subsection{Efficiency Factors}

\begin{figure}
    \centering
    \includegraphics[width=\columnwidth]{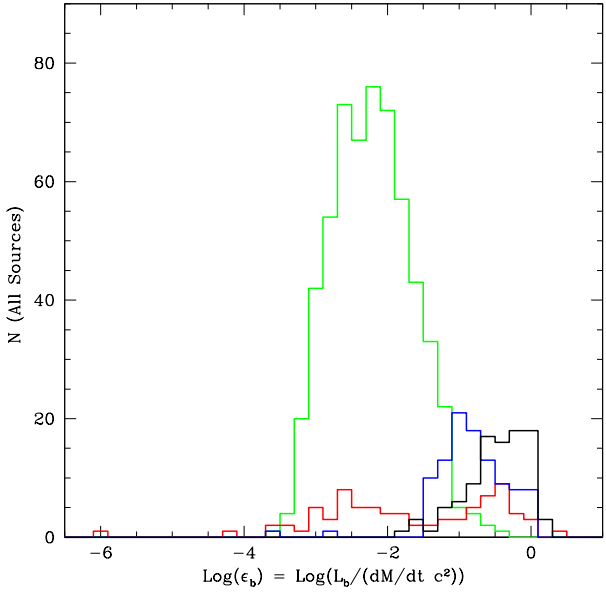}
\caption{Histograms of the bolometric efficiency factor, 
$\epsilon_{bol}$, defined by eq. (9) (see also eq. (5)); sample colors are as in Fig. 
\ref{fig:F1A}.}
\label{fig:F31A}
\end{figure}

\begin{figure}
    \centering
    \includegraphics[width=\columnwidth]{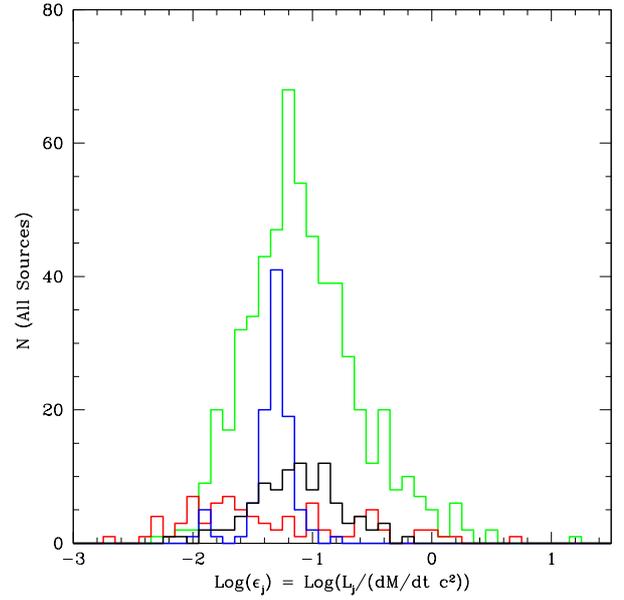}
\caption{Histograms of the outflow efficiency factor, $\epsilon_j$, 
defined in the traditional manner (see eq. 10); sample colors are as in Fig. 
\ref{fig:F1A}.}
\label{fig:F36A}
\end{figure}

\begin{figure}
    \centering
    \includegraphics[width=\columnwidth]{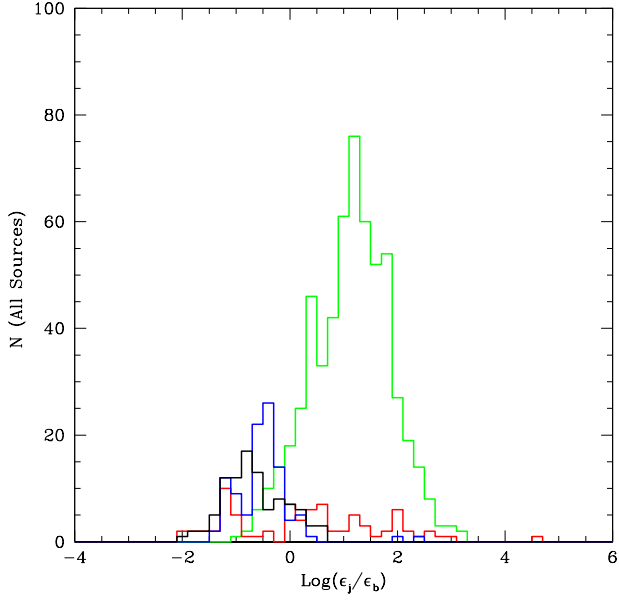}
\caption{Histograms of the ratio of the outflow to the bolometric efficiency factor, 
$\epsilon_j/\epsilon_{bol}$, or $L_j/L_{bol}$; sample colors are as in Fig. 
\ref{fig:F1A}.}
\label{fig:F38A}
\end{figure}

\begin{figure}
    \centering
    \includegraphics[width=\columnwidth]{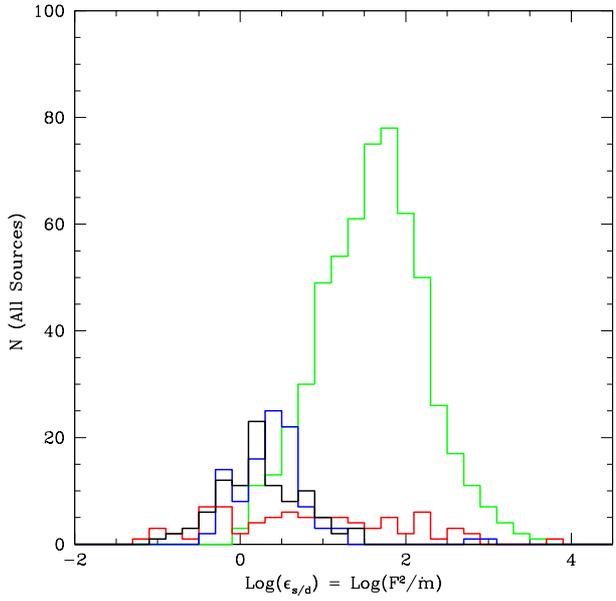}
\caption{Histograms of the "spin relative to disk"  
efficiency factor, $\epsilon_{s/d}$, defined by eq. (11) 
that measures the relative contribution of 
spin and dimensionless mass accretion rate to the Eddington normalized 
beam power, $L_j/L_{Edd}$; sample colors are as in Fig. 
\ref{fig:F1A}.}
\label{fig:F38B}
\end{figure}

Histograms illustrating the efficiency factors 
are shown in Figs. \ref{fig:F31A} - \ref{fig:F38B}. 
The mean value and dispersion for the efficiency factors $\epsilon_{bol}$, 
$\epsilon_j$, the ratio $\epsilon_j/\epsilon_{bol}$, and $\epsilon_{s/d}$, 
defined in section 3.2, are listed in columns (6) - (9) for the samples 
and sources discussed in section 4.1, as described above. 

The efficiency factor for the bolometric luminosity of the accretion disk, 
shown in Fig. \ref{fig:F31A} and listed in column (6) of Table 1 indicates 
a significant overlap of values for the FRII sources and GBH. 
The M03 local sample of AGN appears to split into two components, 
with a break at $\epsilon_{bol} \approx -1.5$. The higher 
efficiency factor component is similar to that of the FRII sources, while 
the lower efficiency factor component is similar to that of the NB16 
LINER sample. To study this further, values of $\epsilon_{bol}$ for each of the 
GBH, and different categories of AGN for the FRII and M03 samples are listed 
in Table 1. The Q, NS1, and S(1-1.9) from the M03 sample overlap with the 
FRII sources, while the L1.9, L2, and S2 M03 sources overlap with the 
NB16. A similar trend is seen for the dimensionless mass accretion 
rates. This is not the case for the mass accretion rates in physical units,
where the NB16 LINERs and M03 sample show significant overlap, while 
the FRII sources and GBH have little overlap with other types of sources. 
The fact that the mean values of $\epsilon_{bol}$ for the samples are quite different, and that 
the dispersion of each of the samples and subsamples is large, indicates 
that a constant value of $\epsilon_{bol}$ does not provide an accurate description 
of these sources. Splitting the samples into sub-sets, as shown in Table 1 
for the GBH, FRII, and M03 sources may provide a way to estimate $\epsilon_{bol}$ for 
each sub-type. Of course, the entire sample of LINERs is one sub-type of AGN. 

The efficiency factor of radiation from the accretion disk, $\epsilon_{bol}$, 
for some sources studied here were obtained by Raimundo et al. (2012) using 
methods that are independent of those developed and applied here. 
Raimundo et al. (2012) used two different methods to obtain bolometric 
disk luminosities, and the resulting efficiency factors are listed 
in Tables 2 and 3 of that paper, referred to here as R12(T2) and R12(T3).  
A comparison of efficiency factors for three sources that are in both R12(T2) 
and R12(T3) indicate that their efficiency factors internally agree to a factor of about 
1.5 - 2. Accretion disk efficiency factors for sources studied here that overlap 
with those studied by Raimundo et al. (2012) are included in columns (6) - (10) 
of the second part of Table 2; there (T5) refers to Table 5 (presented here and included below).
The ratio of values obtained here to those listed in R12(T2) and R12(T3)
are listed in columns (8) and (10) of Table 2, respectively. 
Most of the sources agree to a factor of a few, which is reasonably good 
agreement considering that it is similar to the internal consistency of the comparison sample, 
and considering that the observations used in the 
different studies were obtained at different times, and the sources may 
be intrinsically time variable. 

The beam power efficiency factors are shown in Fig. \ref{fig:F36A} and are 
listed in column (7) of Table 1. These are proportional to $F^2$ (see eq. (10). 
Remarkably, the mean values of $\epsilon_j$ 
for all of the samples and source types are consistent given the dispersion of 
each sample. 
Each AGN sample has a rather broad range of values, 
which is reflected in the large dispersion of each sample. Thus, though 
the mean values of $\epsilon_j$ are similar for all source samples and types
studied, the large dispersion of the AGN samples and sub-types would have to be 
taken into account before assuming and applying a value for similar sources. 
The dispersion of the GBH is relatively small, and the values presented 
here for the GBH could provide a reliable estimate of $\epsilon_j$ (i.e. $F^2$) for 
similar sources.

The ratio $\epsilon_j/\epsilon_{bol}$ (i.e. $L_j/L_{bol}$) is shown in Fig. \ref{fig:F38A} and values 
are listed in column (8) of Table 1. The values for the GBH and FRII sources 
have substantial overlap, and one component of the M03 sources also overlaps 
with these sources. A long tail of values for the M03 sample overlaps with 
the NB16 LINERs. This quantity essentially separates sources into disk 
dominated sources and jet dominated sources. Sources with  
$\epsilon_j/\epsilon_{bol} > 1$ are jet dominated, and those with 
 $\epsilon_j/\epsilon_{bol} < 1$ are disk dominated (see also D18). 
 Clearly, most of the 
 LINERs and some of the M03 sources are jet dominated, most by a very large
 factor of 10 to 100. Outflow from these sources are almost certainly powered 
 by black hole spin, as discussed in section 1.1 (see also section 1.1 of D19). 
 Other sources with very large values of this quantity include A0 6200, and the M03 
 L1.9, L2, and S2 sources. Sources with values less than one include 
 FRII sources of all types, most of the GBH, and the M03 Q, and NS1 sources, 
 while the S(1-1.9) sources have a mean value near 1 and a very large dispersion. 

Another measure of interest is the relative input to the Eddington-normalized 
outflow beam power from black hole spin and from the accretion disk. Eq. (1) 
may be written as $(L_j/L_{Edd}) = g_j (dm/dt) F^2$, 
so one measure of this relative input is given by $\epsilon_{s/d}$, defined by 
eq. (11). Values of this quantity are listed in column (9) of Table 1. 
Almost all of the sources have values of $\epsilon_{s/d}$ that are greater than 
1. Types that include values less than one include about half of the FRII 
quasars (Q), the 3 FRII W sources, and some of the M03 Q and NS1 sources. 
This highlights the role of black hole spin as the source that powers the 
outflow and that of the accretion disk as the source that 
maintains the magnetic field that is anchored in the disk and threads 
the black hole. Note that in the model described by Meier (1999), both 
a disk wind, as in the model of Blandford \& Payne (1982),
and spin energy extraction from the black hole, as in the model of 
Blandford \& Znajek 1977, power the 
outflow, and this model is also described by eq. (1), as discussed in 
detail by D19 section 4 (see also Meier 1999). 

\section{Discussion} 
\label{sec:discussion}
In addition to considering histograms of quantities, it is helpful view the 
redshift distributions of the sources to get a sense of contributions to 
the histograms from sources at different redshift. The GBH are located in 
the Galaxy, so these are not including in the redshift distributions shown in Figs. 
3, 4, and 9-13. The M03 
sources are quite local, and include sources with a wide range of intrinsic 
parameters. The NB16 sources extend to a redshift of about 0.3, with this cutoff 
imposed by NB16 for this study. The D19 sources extend to a redshift of about 2.
For some parameters, it is easy to see the impact of missing lower 
luminosity sources as redshift increases, while other parameters do not 
have a strong redshift dependence. For sources that increase with redshift, 
the upper envelope of the distributions provides a guide as to how parameters 
that describe the most luminous and radio bright sources evolve with redshift. 

It is clear from Fig. \ref{fig:F4A} that sources at higher redshift have higher 
mass accretion rates  physical units, so the high mass 
accretion rates seen in Fig. \ref{fig:F3A} are due
to these sources. The low redshift sample of M03 suggests that there are sources
at all accretion rates, and that the lower luminosity sources that are associated 
with lower mass accretion rates are missing at higher redshift due to the flux 
limited nature of the samples that include higher redshift sources. 
Fig. \ref{fig:F2A} 
shows the dimensionless mass accretion rate as a function of redshift. 
Clearly, there is a rather sharp boundary at $dm/dt = 1$, and this is not 
artificially imposed, but is an empirical result and reflects that very sharp 
boundary found by D16, D18, and D19 when the bolometric 
disk luminosity approaches the Eddington luminosity. While 
$dM/dt \propto M_{dyn}(dm/dt)$, the accretion disk magnetic field 
strength $B$ in physical units is obtained from $B^2 \propto M_{dyn}^{-1} (dm/dt)$, 
which accounts for the redshift distribution of FRII sources seen in Fig. 
\ref{fig:F6A}. This suggests a weak dependence of the disk magnetic field 
strength, $B$, in physical units on redshift. 

\begin{figure}
    \centering
    \includegraphics[width=\columnwidth]{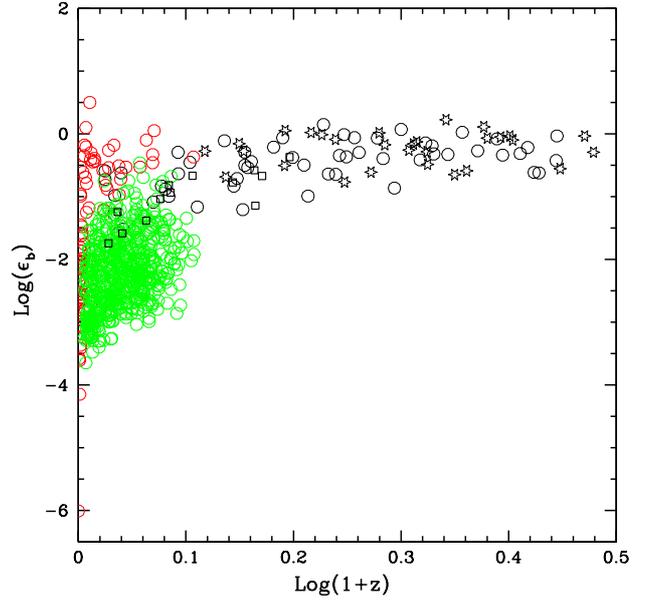}
\caption{The bolometric efficiency factor, $\epsilon_{bol}$,
is shown as a function of $(1+z)$ for the AGN samples; 
colors and symbols are as in Fig. \ref{fig:F2A}.}
\label{fig:F34A}
\end{figure}

\begin{figure}
    \centering
    \includegraphics[width=\columnwidth]{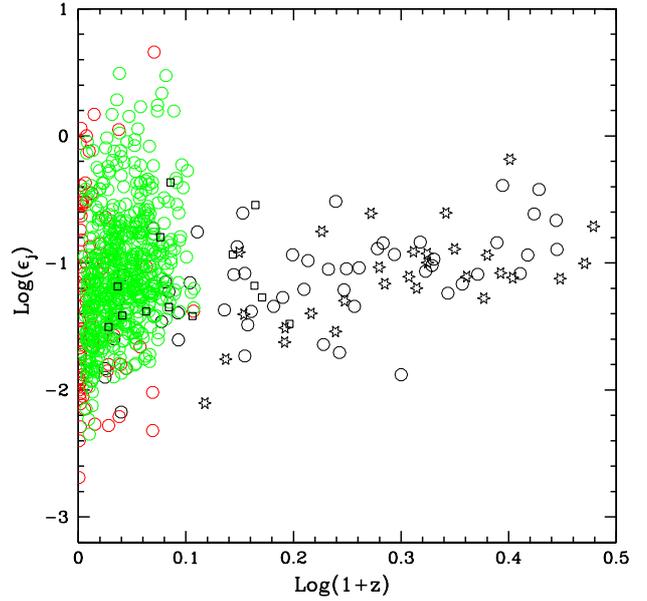}
\caption{The outflow efficiency factor, $\epsilon_j$,
is shown as a function of $(1+z)$ for the AGN samples; 
colors and symbols are as in Fig. \ref{fig:F2A}.}
\label{fig:F33A}
\end{figure}

\begin{figure}
    \centering
    \includegraphics[width=\columnwidth]{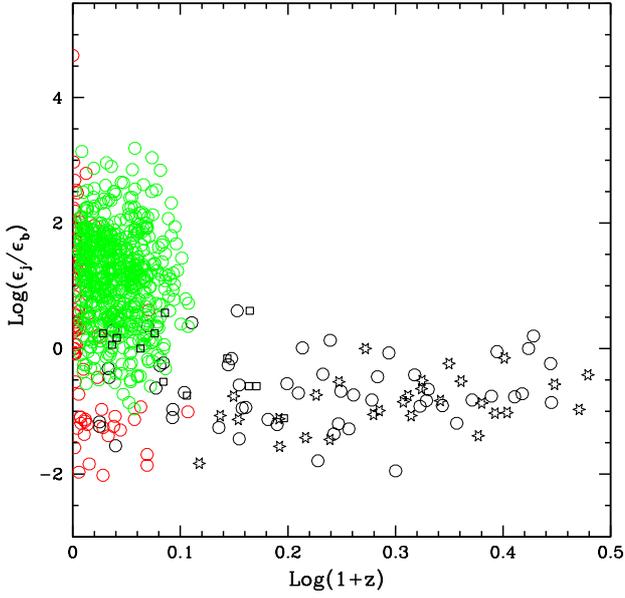}
\caption{The ratio of the outflow to the bolometric efficiency factor, 
$\epsilon_j/\epsilon_{bol} = L_j/L_{bol}$, is shown as a function of $(1+z)$ for the AGN samples; 
colors and symbols are as in Fig. \ref{fig:F2A}.}
\label{fig:F35A}
\end{figure}

\begin{figure}
    \centering
    \includegraphics[width=\columnwidth]{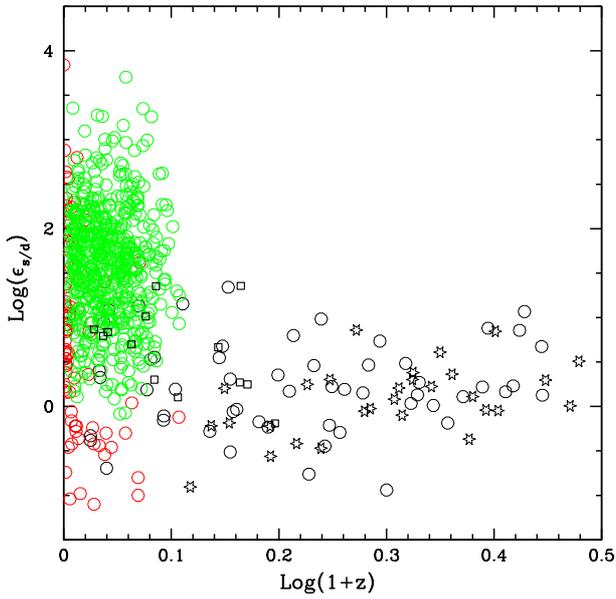}
\caption{The spin relative to disk efficiency factor, $\epsilon_{s/d}$, 
is shown as a function of $(1+z)$ for the AGN samples; 
colors and symbols are as in Fig. \ref{fig:F2A}.}
\label{fig:F35C}
\end{figure}

\begin{figure}
    \centering
    \includegraphics[width=\columnwidth]{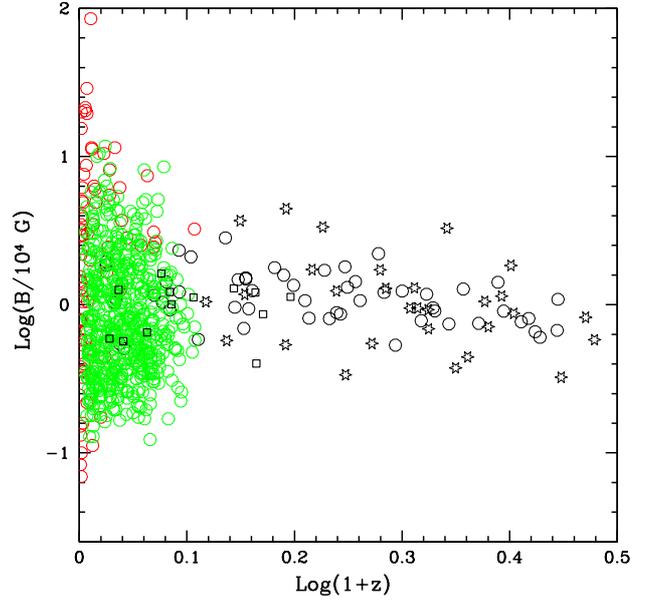}
\caption{The accretion disk magnetic field strength at the location where 
the field that taps the black hole spin is anchored in the disk in units of 
$10^4$ G as a function of $(1+z)$ for the AGN samples; 
colors and symbols are as in Fig. \ref{fig:F2A}.}
		  \label{fig:F6A}
\end{figure}

The efficiency factors are shown as a function of redshift in Figs. 
\ref{fig:F34A} to \ref{fig:F35C}. Fig. \ref{fig:F34A} reflects the fact 
that there are no super-Eddington sources. There is a narrow range of
values for the FRII sources with little or no redshift evolution. 
It is clear that NB16 LINERs with lower values 
of $\epsilon_{bol}$ that are present near zero redshift have dropped out by 
a redshift of about 0.3, most likely reflecting a selection effect whereby 
lower luminosity sources are not included in the sample 
as redshift increases (NB16). Redshift 
selection effects for $\epsilon_j$ and $\epsilon_j/\epsilon_{bol}$ for each sample 
separately are much less pronounced, especially that for 
$\epsilon_j/\epsilon_{bol}$, compared with that for $\epsilon_{bol}$. 

\section{Summary \& Conclusions}
\label{sec:conclusions}

The outflow method allows empirical 
estimates of the accretion disk magnetic field strength in 
Eddington units, $(B/B_{Edd})$, at the location where the magnetic field 
that torques the black hole is anchored in the disk (D19). 
Interestingly, two key physical parameters describing the 
accretion disk can be obtained from this ratio: the magnetic 
field strength in physical units, and the 
mass accretion rate in physical units. As shown in sections 1.1 and 3.1,  
the mass accretion rate in physical units is $dM/dt \propto L_{Edd} (B/B_{Edd})^2$, 
whereas the accretion disk magnetic field strength in physical units is 
$B^2 \propto L_{Edd}^{-1} (B/B_{Edd})^2$ (D19). So, three important physical 
quantities that describe 
the accretion disk follow from empirical estimates of $(B/B_{Edd})^2$: 
the dimensionless mass accretion rate; the mass accretion rate 
in physical units; and the disk magnetic 
field strength in physical units (disk magnetic pressure in physical 
units). Here, data for three samples of AGN, including 756 supermassive black 
hole systems, and 103 measurements of 4 stellar mass black holes, 
referred to as GBH, are used to obtain and study mass accretion rates and efficiency factors, while
the magnetic field strength in physical and dimensionless units are discussed by D19. 

Mass accretion rates in dimensionless units can be divided into five categories, 
defined here as high (${\rm{Log}}(\dot{m}) \geq -0.5)$, 
medium-high ($-0.5 > {\rm{Log}}(\dot{m}) \geq -1.0)$, 
medium ($-1.0 > {\rm{Log}}(\dot{m}) \geq -1.5)$, 
medium-low ($-1.5 > {\rm{Log}}(\dot{m}) \geq -2.0)$, 
and low $({\rm{Log}}(\dot{m}) < -2.0)$. Broadly, sources fall into the following 
categories: all FRII sources (except LEG), and M03 NS1 and Q sources 
have high ${\rm{Log}}(\dot{m})$; three GBH (all except A06200), the D16/19 
FRII LEG, and the M03 S(1-1.9) sources have medium-high ${\rm{Log}}(\dot{m})$;
the M03 S2 sources have medium ${\rm{Log}}(\dot{m})$; GBH A0 6200, 
the NB16 LINERs, 
and M03 L1.9 and L2 sources have medium-low ${\rm{Log}}(\dot{m})$; a 
few individual sources have low ${\rm{Log}}(\dot{m})$ (see Tables 3-6). 

Mass accretion rates in physical units for the AGN studied here 
seem to have distinct mean values, but with a broad dispersion, as is
evident in Fig. 2, and appear to fall 
into four broad categories: high, medium, low, and very low. 
Sources with high mass 
accretion rates, where $dM/dt$ is in units of $M_{\odot}/yr$, 
have ($\rm{Log}(dM/dt) > 0$) and include FRII Q and some HEG; sources with medium 
rates of about $(0 > \rm{Log}(dM/dt) > -1)$ include FRII LEG, W, and 
some HEG types and M03 Q types; sources with mean values in the low range 
from $(-1 > \rm{Log}(dM/dt) > -2)$ are practically absent; 
and sources with very low rates of about ($\rm{Log}(dM/dt) < -2$) 
include the NB16 LINERs and all of the M03 sources except Q type. 
Though sources with mean values in the low range are 
practically absent, many individual sources fall into this range 
and are evident in the dispersion of the mean values of $dM/dt$.
The GBH all have very similar mass accretion rates of about $\rm{Log}(dM/dt) 
\approx -8.5$, with a few individual observations stretching to very low 
values of about $(10^{-10} - 10^{-11})$, for example one observation of 
A0 6200 and one of V404 Cyg (see Table 3). 

Efficiency factors associated with the bolometric luminosity of the 
accretion disk have a broad range, and may be categorized as high 
$(\rm{Log}(\epsilon_{bol}) > -1)$, medium $(-1 > \rm{Log}(\epsilon_{bol}) > -2)$, 
and low $(\rm{Log}(\epsilon_{bol}) < -2)$. Sources with high values include
most or all of the FRII sources (except about half of the FRII LEG, which fall 
into the medium category), GX 339-4, about half of the observations of 
V404 Cyg, and the M03 NS1 and Q sources. Sources in the medium category include 
about half of the FRII LEG sources and about half of the V404 Cyg observations, 
J1118+480, and the M03 S(1-1.9) and S2 types. Sources in the low category include 
the NB16 LINERs, A0 6200, and the M03 L1.9 and L2 types. 

The mean values of efficiency factors associated with collimated outflows, 
$\epsilon_j$, 
defined in the traditional manner in terms of mass accretion rate, have 
a small range of mean values and a large dispersion per sample (see 
the first four lines of Table 1 and Fig. 6). 
This efficiency factor measures the beam power relative to the mass accretion 
rate in physical units. In the spin powered outflow model described in section 1.1, 
the accretion disk plays the role of maintaining the magnetic field; 
the field is anchored in the disk and threads the black hole, exerting a
torque and producing the outflow (Blandford \& Znajek 1977), while it is 
the spin energy of the hole that powers the outflow. 
In the Meier model (1999), the outflow is powered by both the spin of 
the black hole and a disk wind, with the disk wind component similar to that 
of Blandford \& Payne model (1982). The Meier model has the same functional 
form as the Blandford \& Znajek model, but has a different constant of 
proportionality, and thus is described by eq. (1), as discussed in detail by D19.  

A measure of the outflow power sourced to the black hole spin relative 
to that contributed by the disk is the efficiency factor $\epsilon_{s/d}$, 
listed in column (9) of Table 1. All of the samples, and most of the sub-samples,
have mean values of $\rm{Log}(\epsilon_{s/d}) > 0$, indicating that most of the 
outflow beam power originates from the black hole spin. The exceptions are
the 3 FRII W sources, and a subset of the 20 M03 NS1 and Q sources. 

The ratio $\epsilon_j/\epsilon_{bol} = L_j/L_{bol}$ was studied in detail 
by D18. As discussed in that work, sources with $\epsilon_j/\epsilon_{bol} > 1$ 
are jet dominated sources, while those with $\epsilon_j/\epsilon_{bol} < 1$ 
are disk dominated sources. The results obtained by D18 align quite well 
with previously published results, such as those discussed by Heckman \& Best 
(2014) and NB16. The results obtained here indicate that both disk-dominated and 
jet-dominated sources have outflow beam powers with a spin term that exceeds 
the disk term, indicated by a value of $(\epsilon_{s/d}) > 1$. 
That is, almost all of the sources with $\epsilon_j/\epsilon_{bol} > 1$ have 
$(\epsilon_{s/d}) > 1$, and many sources with $\epsilon_j/\epsilon_{bol} < 1$ have 
$(\epsilon_{s/d}) > 1$ including three of the GBH, all of the FRII sources 
except the W types, and almost all of the M03 types except NS1 and Q.  

Parameters with a dispersion that is not too large could provide estimates 
of this parameter for additional sources of the same type. If the parameter 
evolves with redshift, then this should only be applied to sources of the same 
type within a similar redshift range. To identify such source types, consider 
quantities with a dispersion less than about 0.5 dex to one significant figure. 
This suggests that the quantities that could be applied to other sources of 
the same type are: 
for NB16 LINERs, ${\rm{Log}}(\dot{m})$, $\rm{Log}(dM/dt)$, and $\epsilon_j$;
for FRII sources, ${\rm{Log}}(\dot{m})$, $\epsilon_{bol}$, 
$\epsilon_j$, and $\epsilon_{s/d}$; 
for M03 sources, none for the full sample but may be applied to sub-samples; for the GBH, 
${\rm{Log}}(\dot{m})$, $\rm{Log}(dM/dt)$, $\epsilon_{bol}$, 
$\epsilon_j$, and $\epsilon_{s/d}$. 
The results obtained here suggest that 
black hole systems with powerful outflows provide an interesting and 
important guide for mass accretion rates and efficiency factors that 
may be applied to additional sources for some source types. 

\section*{Data Availability Statement}
The data underlying this article are available in the article or 
are listed in D19. 

\section*{Acknowledgments}
Thanks are extended to the referee for very helpful comments and suggestions.
It is a pleasure to thank the many colleagues with whom this work was discussed, 
especially Jean Brodie,  George Djorgovski, Megan Donahue, Yan-Fei Jiang, 
Syd Meshkov, Chiara Mingarelli, Chris O'Dea, Masha Okounkova, David Spergel, 
Alan Weinstein, and Rosie Wyse. 
I would especially like to thank Jerry Ostriker for helpful discussions 
related to the topics discussed in this paper. The Flatiron Institute is 
supported by the Simons Foundation. This work was also supported in part by 
the Losoncy Fund and the PSU Berks Advisory Board,  
and was performed in part at the Aspen Center for Physics, 
which is supported by National Science Foundation grant PHY-1607611. 

\clearpage
\begin{landscape}
\begin{table}
\begin{minipage}{165mm}
\caption{Mean Value and Dispersion of Mass Accretion Rates and Efficiency Factors for each Sample, 
Sub-Sample, or Source; the uncertainty per 
source of each quantity is indicated in (brackets) in the top part of the Table.}   
\label{tab:1}        
\begin{tabular}{llllllllll}   
\hline\hline                    
(1)&(2)&(3)&(4)&(5)&(6)&(7)&(8)&(9)&(10)\\
\hline 
&&	&$\rm{Log}$&$\rm{Log}$&$\rm{Log}$&$\rm{Log}$&$	\rm{Log}$&$	\rm{Log}$&$	\rm{Log}$	\\
Sample&type&N		&	$(dm/dt)	$&	$(dM/dt)$&	($\epsilon_{bol}$)	&	($\epsilon_j$)	&	
($\epsilon_j/\epsilon_{bol}$)	& ($\epsilon_{s/d})$	&	($M_{dyn}$)\\
or Source&&&&$(M_{\odot}/yr)	$&&&&&$	(M_{\odot})	$\\				
\hline				
NB16&AGN	&$	576	$&$	-1.67	\pm	0.42(0.20)	$&$	-2.24	\pm	0.51(0.23)	$&$	-2.21	\pm	0.56(0.24)	$&$	-1.07	\pm	0.47(0.28)	$&$	1.14	\pm	0.73(0.31)	$&$	1.60	\pm	0.64(0.29)	$&$	8.07	\pm	0.50 (0.35)	$\\									
D16/D19	&AGN&$	100	$&$	-0.38	\pm	0.34(0.24)	$&$	0.09	\pm	0.60(0.26)	$&$	-0.48	\pm	0.43(0.29)	$&$	-1.15	\pm	0.38(0.29)	$&$	-0.67	\pm	0.57(0.40)	$&$	0.23	\pm	0.51(0.37)	$&$	9.11	\pm	0.41 (0.35)	$\\			
M03&AGN	&$	80	$&$	-1.17	\pm	0.84(0.26)	$&$	-1.97	\pm	0.99(0.30)	$&$	-1.68	\pm	1.22(0.32)	$&$	-1.33	\pm	0.71(0.38)	$&$	0.35	\pm	1.40(0.36)	$&$	0.83	\pm	1.10(0.35)	$&$	7.84	\pm	0.81 (0.50)	$\\						
S15	&GBH&$	103	$&$	-0.74	\pm	0.46(0.12)	$&$	-8.56	\pm	0.45(0.12)	$&$	-0.84	\pm	0.52(0.15)	$&$	-1.34	\pm	0.18(0.14)	$&$	-0.51	\pm	0.56(0.28)	$&$	0.40	\pm	0.50(0.25)	$&$	0.83	\pm	0.09	$\\					
\hline					
GX 339-4	&GBH&$	76	$&$	-0.65	\pm	0.36	$&$	-8.52	\pm	0.36	$&$	-0.74	\pm	0.40	$&$	-1.35	\pm	0.13	$&$	-0.61	\pm	0.41	$&$	0.31	\pm	0.37	$&$	0.78			$\\			V404 Cyg	&GBH&$	20	$&$	-0.88	\pm	0.47	$&$	-8.53	\pm	0.47	$&$	-0.99	\pm	0.53	$&$	-1.20	\pm	0.12	$&$	-0.21	\pm	0.62	$&$	0.68	\pm	0.56	$&$	1.00			$\\						
J1118+480	&GBH&$	5	$&$	-1.08	\pm	0.02	$&$	-8.85	\pm	0.02	$&$	-1.22	\pm	0.04	$&$	-1.86	\pm	0.02	$&$	-0.63	\pm	0.06	$&$	0.23	\pm	0.04	$&$	0.88			$\\			
A0 6200	&GBH&$	2	$&$	-1.85	\pm	1.97	$&$	-9.68	\pm	1.97	$&$	-2.08	\pm	2.23	$&$	-1.21	\pm	0.05	$&$	0.88	\pm	2.28	$&$	1.66	\pm	2.01	$&$	0.82			$\\						\hline				
FRII &W&$	3	$&$	-0.16	\pm	0.31	$&$	-0.28	\pm	0.19	$&$	-0.20	\pm	0.39	$&$	-1.41	\pm	0.26	$&$	-1.21	\pm	0.36	$&$	-0.25	\pm	0.30	$&$	8.52	\pm	0.24 (0.35)	$\\										
FRII &Q&$	29	$&$	-0.18	\pm	0.21	$&$	0.53	\pm	0.38	$&$	-0.23	\pm	0.27	$&$	-1.12	\pm	0.38	$&$	-0.88	\pm	0.42	$&$	0.07	\pm	0.40	$&$	9.36	\pm	0.46 (0.40)	$\\										
FRII &HEG&$	55	$&$	-0.40	\pm	0.31	$&$	0.03	\pm	0.54	$&$	-0.51	\pm	0.40	$&$	-1.16	\pm	0.38	$&$	-0.65	\pm	0.58	$&$	0.24	\pm	0.52	$&$	9.07	\pm	0.33 (0.35)	$\\										
FRII &LEG&$	13	$&$	-0.78	\pm	0.32	$&$	-0.58	\pm	0.52	$&$	-0.99	\pm	0.41	$&$	-1.14	\pm	0.37	$&$	-0.15	\pm	0.53	$&$	0.64	\pm	0.47	$&$	8.84	\pm	0.31 (0.35)	$\\	
\hline					
M03 &NS1&$	7	$&$	-0.24	\pm	0.31	$&$	-1.99	\pm	0.63	$&$	-0.35	\pm	0.45	$&$	-1.35	\pm	0.49	$&$	-1.00	\pm	0.61	$&$	-0.11	\pm	0.54	$&$	6.90	\pm	0.63 (0.50)	$\\										
M03 &Q&$	13	$&$	-0.33	\pm	0.20	$&$	-0.85	\pm	0.46	$&$	-0.47	\pm	0.29	$&$	-1.50	\pm	0.91	$&$	-1.02	\pm	0.79	$&$	-0.17	\pm	0.82	$&$	8.13	\pm	0.42 (0.50)	$\\										
M03 &S(1-1.9)&$	17	$&$	-0.89	\pm	0.66	$&$	-2.03	\pm	0.71	$&$	-1.27	\pm	0.95	$&$	-1.30	\pm	0.76	$&$	-0.03	\pm	1.07	$&$	0.58	\pm	0.89	$&$	7.50	\pm	0.46 (0.50)	$\\										
M03 &S2&$	22	$&$	-1.27	\pm	0.54	$&$	-2.17	\pm	0.96	$&$	-1.83	\pm	0.77	$&$	-1.20	\pm	0.70	$&$	0.63	\pm	1.09	$&$	1.07	\pm	0.92	$&$	7.75	\pm	0.90 (0.50)	$\\	
M03 &L1.9&$	10	$&$	-1.92	\pm	0.36	$&$	-2.19	\pm	0.71	$&$	-2.75	\pm	0.52	$&$	-1.34	\pm	0.66	$&$	1.41	\pm	0.85	$&$	1.57	\pm	0.76	$&$	8.37	\pm	0.66 (0.50)	$\\										
M03 &L2&$	10	$&$	-2.09	\pm	0.40	$&$	-2.26	\pm	0.65	$&$	-3.00	\pm	0.58	$&$	-1.42	\pm	0.74	$&$	1.58	\pm	1.02	$&$	1.66	\pm	0.90	$&$	8.47	\pm	0.75 (0.50)	$\\		
\hline					
\hline
\end{tabular}
\end{minipage}
\end{table}
\end{landscape}
\clearpage

\begin{table*}
\begin{minipage}{165mm}
\caption{Comparison of Accretion Rates and Bolometric Efficiency Factors 
Obtained Here with Independently Determined Values (see section 4).}   
\label{tab:2}        
\begin{tabular}{llllllllll}   
\hline\hline                    
(1)&(2)&(3)&(4)&(5)\\
Source	&Type		&J19&This Work (T5)& Ratio\\
		&		&$(dm/dt)$&$(dm/dt)$&(T5)/J19\\
\hline
Ark	$	564	$&	NS1	&$	1.70	$&$	2.21	$&$	1.30											$\\
Mrk	$	279	$&	S1.5	&$	0.75	$&$	0.48	$&$	0.64											$\\
Mrk	$	335	$&	NS1	&$	0.74	$&$	0.51	$&$	0.69											$\\
Mrk	$	590	$&	S1.2	&$	0.31	$&$	0.61	$&$	1.97											$\\
PG	$	0804+761	$&	Q	&$	1.13	$&$	0.43	$&$	0.38											$\\
PG	$	0844+349	$&	Q	&$	1.20	$&$	0.42	$&$	0.35											$\\
PG	$	1229+204	$&	Q	&$	0.50	$&$	0.65	$&$	1.29											$\\
PG	$	1426+015	$&	Q	&$	0.28	$&$	0.21	$&$	0.75											$\\																\hline	
\hline
(1)&(2)&(3)&(4)&(5)&(6)&(7)&(8)&(9)&(10)\\
Source	&Type		&R12 (T1)&This Work (T5)&Ratio&R12 (T3)&This Work (T5)&Ratio
&R12 (T2)&Ratio\\
		&		&$(dM/dt)$&$(dM/dt)$& (T5)/R12(T1)&
			$\epsilon_{bol}$&$\epsilon_{bol}$&(T5)/R12(T3)&$\epsilon_{bol}$&(T5)/R12(T2)\\
				&		&$(M_{\odot} ~yr^{-1})$&$(M_{\odot}~ yr^{-1})$&
			\\
\hline	
Mrk	$	279	$&	S1.5	&$	0.045	$&$	0.045	$&$	1.00	$&$	0.25	$&$	0.35	$&$	1.40	$&$		$&$		$\\
Mrk	$	509	$&	NS1	&$	0.095	$&$	0.082	$&$	0.86	$&$	0.29	$&$	0.37	$&$	1.28	$&$	0.46	$&$	0.80	$\\
NGC	$	5548	$&	S1.5	&$	0.006	$&$	0.067	$&$	11.64	$&$	0.98	$&$	0.16	$&$	0.16	$&$		$&$		$\\
NGC	$	7469	$&	S1	&$	0.083	$&$	0.010	$&$	0.12	$&$	0.21	$&$	0.60	$&$	2.86	$&$	0.26	$&$	2.31	$\\
3C	$	120	$&	S1	&$	0.178	$&$	0.065	$&$	0.37	$&$	0.16	$&$	0.42	$&$	2.63	$&$		$&$		$\\
3C	$	390.3	$&	S1	&$	0.062	$&$	0.202	$&$	3.28	$&$	0.36	$&$	0.15	$&$	0.42	$\\
\hline
\hline 
\end{tabular}
\end{minipage}
\end{table*}

\clearpage

\begin{table*}
\begin{minipage}{165mm}
\scriptsize
\caption{Mass Accretion Rates and Efficiency Factors for Stellar-Mass 
Galactic Black Holes in X-ray Binary Systems.}   
\label{tab:3}        
\begin{tabular}{llllllll}   
\hline\hline                    
(1)&(2)&(3)&(4)&(5)&(6)&(7)&(8)\\
	&$\rm{Log}$&$\rm{Log}$&$\rm{Log}$&$\rm{Log}$&$	\rm{Log}$&$	\rm{Log}$&$	\rm{Log}$	\\
Source		&	$(dm/dt)	$&	$(dM/dt)$&	($\epsilon_{bol}$)	&	($\epsilon_j$)	&	
($\epsilon_j/\epsilon_{bol}$)	&	($\epsilon_{s/d}$)	&	($M_{dyn}$)\\
&&$(M_{\odot}/yr)	$&&&&&$	(M_{\odot})	$\\
\hline
GX	$	339-4	$&$	-0.38	$&$	-8.25	$&$	-0.45	$&$	-1.43	$&$	-0.98	$&$	-0.04	$&$	0.78	$\\
GX	$	339-4	$&$	-0.38	$&$	-8.25	$&$	-0.44	$&$	-1.43	$&$	-0.99	$&$	-0.04	$&$	0.78	$\\
GX	$	339-4	$&$	-0.58	$&$	-8.45	$&$	-0.67	$&$	-1.29	$&$	-0.62	$&$	0.30	$&$	0.78	$\\
GX	$	339-4	$&$	-0.66	$&$	-8.53	$&$	-0.72	$&$	-1.21	$&$	-0.49	$&$	0.44	$&$	0.78	$\\
\hline
\end{tabular}
\end{minipage}
\end{table*}

\begin{table*}
\begin{minipage}{165mm}
\scriptsize
\contcaption{Mass Accretion Rates and Efficiency Factors for 
Stellar-Mass Galactic Black Holes in X-ray Binary Systems.}   
\label{tab:3A}        
\begin{tabular}{llllllll}   
\hline\hline                    
(1)&(2)&(3)&(4)&(5)&(6)&(7)&(8)\\
	&$\rm{Log}$&$\rm{Log}$&$\rm{Log}$&$\rm{Log}$&$	\rm{Log}$&$	\rm{Log}$&$	\rm{Log}$	\\
Source		&	$(dm/dt)	$&	$(dM/dt)$&	($\epsilon_{bol}$)	&	($\epsilon_j$)	&	
($\epsilon_j/\epsilon_{bol}$)	&	($\epsilon_{s/d}$)	&	($M_{dyn}$)\\
&&$(M_{\odot}/yr)	$&&&&&$	(M_{\odot})	$\\
\hline
XTE	$	J1118+480	$&$	-1.06	$&$	-8.83	$&$	-1.18	$&$	-1.88	$&$	-0.70	$&$	0.18	$&$	0.88	$\\
XTE	$	J1118+480	$&$	-1.10	$&$	-8.87	$&$	-1.25	$&$	-1.84	$&$	-0.59	$&$	0.26	$&$	0.88	$\\
XTE	$	J1118+480	$&$	-1.10	$&$	-8.87	$&$	-1.24	$&$	-1.84	$&$	-0.60	$&$	0.26	$&$	0.88	$\\
XTE	$	J1118+480	$&$	-1.10	$&$	-8.87	$&$	-1.26	$&$	-1.84	$&$	-0.58	$&$	0.26	$&$	0.88	$\\
\hline					
\hline
\end{tabular}
\end{minipage}
\end{table*}

\begin{table*}
\begin{minipage}{165mm}
\scriptsize
\caption{Mass Accretion Rates and Efficiency Factors for the Classical Double (FRII)  Sources.}  
\label{tab:4A}        
\begin{tabular}{llllllllll}   
\hline\hline                    
(1)&(2)&(3)&(4)&(5)&(6)&(7)&(8)&(9)&(10)\\
	&		&	&$\rm{Log}$&$\rm{Log}$&$\rm{Log}$&$\rm{Log}$&$	\rm{Log}$&$	\rm{Log}$&$	\rm{Log}$	\\
Source		&		type	&	z	&	$(dm/dt)	$&	$(dM/dt)$&	($\epsilon_{bol}$)	&	($\epsilon_j$)	&	
($\epsilon_j/\epsilon_{bol}$)	&	($\epsilon_{s/d}$)	&	($M_{dyn}$)\\
&&&&$(M_{\odot}/yr)	$&&&&&$	(M_{\odot})	$\\
\hline																3C	33	&	HEG	&$	0.059	$&$	-0.46	$&$	-0.54	$&$	-0.58	$&$	-1.83	$&$	-1.25	$&$	-0.37	$&$	8.56	$\\
3C	192	&	HEG	&$	0.059	$&$	-0.57	$&$	-0.81	$&$	-0.73	$&$	-1.90	$&$	-1.17	$&$	-0.33	$&$	8.41	$\\
3C	285	&	HEG	&$	0.079	$&$	-1.00	$&$	-1.12	$&$	-1.28	$&$	-1.60	$&$	-0.32	$&$	0.40	$&$	8.53	$\\
3C	452	&	HEG	&$	0.081	$&$	-0.77	$&$	-0.70	$&$	-0.98	$&$	-1.44	$&$	-0.47	$&$	0.32	$&$	8.71	$\\
\hline
\end{tabular}
\end{minipage}
\end{table*}

\begin{table*}
\begin{minipage}{165mm}
\scriptsize
\contcaption{Mass Accretion Rates and Efficiency Factors for the Classical Double (FRII)  Sources.}  
\label{tab:4B}        
\begin{tabular}{llllllllll}   
\hline\hline                    
(1)&(2)&(3)&(4)&(5)&(6)&(7)&(8)&(9)&(10)\\
	&		&	&$\rm{Log}$&$\rm{Log}$&$\rm{Log}$&$\rm{Log}$&$	\rm{Log}$&$	\rm{Log}$&$	\rm{Log}$	\\
Source		&		type	&	z	&	$(dm/dt)	$&	$(dM/dt)$&	($\epsilon_{bol}$)	&	($\epsilon_j$)	&	
($\epsilon_j/\epsilon_{bol}$)	&	($\epsilon_{s/d}$)	&	($M_{dyn}$)\\
&&&&$(M_{\odot}/yr)	$&&&&&$	(M_{\odot})	$\\
\hline		
3C	35	&	LEG	&$	0.067	$&$	-1.37	$&$	-1.36	$&$	-1.74	$&$	-1.51	$&$	0.23	$&$	0.86	$&$	8.65	$\\
3C	326	&	LEG	&$	0.088	$&$	-0.97	$&$	-1.24	$&$	-1.24	$&$	-1.18	$&$	0.06	$&$	0.79	$&$	8.38	$\\
3C	236	&	LEG	&$	0.099	$&$	-1.24	$&$	-1.07	$&$	-1.57	$&$	-1.41	$&$	0.17	$&$	0.83	$&$	8.81	$\\
4C	12.03	&	LEG	&$	0.156	$&$	-1.08	$&$	-0.87	$&$	-1.37	$&$	-1.37	$&$	0.00	$&$	0.70	$&$	8.85	$\\
\hline					
\hline
\end{tabular}
\end{minipage}
\end{table*}

\begin{table*}
\begin{minipage}{165mm}
\scriptsize
\caption{Mass Accretion Rates and Efficiency Factors for the M03 Local Sample of AGN.}  
\label{tab:5A}        
\begin{tabular}{llllllllll}   
\hline\hline                    
(1)&(2)&(3)&(4)&(5)&(6)&(7)&(8)&(9)&(10)\\
	&		&	&$\rm{Log}$&$\rm{Log}$&$\rm{Log}$&$\rm{Log}$&$	\rm{Log}$&$	\rm{Log}$&$	\rm{Log}$	\\
Source		&		type	&	D(Mpc)	&	$(dm/dt)	$&	$(dM/dt)$&	($\epsilon_{bol}$)	&	($\epsilon_j$)	&	
($\epsilon_j/\epsilon_{bol}$)	&	($\epsilon_{s/d}$)	&	($M_{dyn}$)\\
&&&&$(M_{\odot}/yr)	$&&&&&$	(M_{\odot})	$\\
\hline				
Ark	$	564	$&	NS1	&$	106	$&$	0.34	$&$	-2.27	$&$	0.50	$&$	-0.87	$&$	-1.37	$&$	-0.22	$&$	6.03	$\\
Cyg	$	A	$&	S2/L2	&$	240	$&$	-0.84	$&$	-0.09	$&$	-1.21	$&$	-1.00	$&$	0.21	$&$	0.84	$&$	9.40	$\\
IC	$	1459	$&	L2	&$	31	$&$	-2.14	$&$	-1.79	$&$	-3.07	$&$	-0.55	$&$	2.52	$&$	2.58	$&$	9.00	$\\
IC	$	4296	$&	L1.9	&$	201	$&$	-1.96	$&$	-1.51	$&$	-2.81	$&$	-0.85	$&$	1.96	$&$	2.10	$&$	9.10	$\\
\hline
\end{tabular}
\end{minipage}
\end{table*}

\begin{table*}
\begin{minipage}{165mm}
\scriptsize
\contcaption{Mass Accretion Rates and Efficiency Factors for the M03 Local Sample of AGN.}  
\begin{tabular}{llllllllll}   
\hline\hline                    
(1)&(2)&(3)&(4)&(5)&(6)&(7)&(8)&(9)&(10)\\
	&		&	&$\rm{Log}$&$\rm{Log}$&$\rm{Log}$&$\rm{Log}$&$	\rm{Log}$&$	\rm{Log}$&$	\rm{Log}$	\\
Source		&		type	&	D(Mpc)	&	$(dm/dt)	$&	$(dM/dt)$&	($\epsilon_{bol}$)	&	($\epsilon_j$)	&	
($\epsilon_j/\epsilon_{bol}$)	&	($\epsilon_{s/d}$)	&	($M_{dyn}$)\\
&&&&$(M_{\odot}/yr)	$&&&&&$	(M_{\odot})	$\\
\hline				
PG	$	1613+658	$&	Q	&$	605	$&$	-0.37	$&$	-0.64	$&$	-0.54	$&$	-1.67	$&$	-1.13	$&$	-0.30	$&$	8.38	$\\
PG	$	2130+099	$&	Q	&$	273	$&$	-0.61	$&$	-1.10	$&$	-0.87	$&$	-1.84	$&$	-0.97	$&$	-0.23	$&$	8.16	$\\
3C	$	120	$&	S1	&$	148	$&$	-0.27	$&$	-1.18	$&$	-0.38	$&$	0.18	$&$	0.56	$&$	1.44	$&$	7.73	$\\
3C	$	390.3	$&	S1	&$	258	$&$	-0.58	$&$	-0.69	$&$	-0.82	$&$	-0.64	$&$	0.18	$&$	0.93	$&$	8.53	$\\
Sgr	$	A^*	$&		&$	0.008	$&$	-4.18	$&$	-6.23	$&$	-6.01	$&$	-1.34	$&$	4.67	$&$	3.84	$&$	6.60	$\\			
\hline					
\hline
\end{tabular}
\end{minipage}
\end{table*}

\clearpage

\begin{table*}
\begin{minipage}{165mm}
\caption{Mass Accretion Rates and Efficiency Factors for the NB16 LINERs.}   
\label{tab:6A}        
\begin{tabular}{lllllllll}   
\hline\hline                    
(1)&(2)&(3)&(4)&(5)&(6)&(7)&(8)&(9)\\
NB16&&$\rm{Log}$	&$\rm{Log}$&$\rm{Log}$&$\rm{Log}$&$\rm{Log}$&$	\rm{Log}$&$	\rm{Log}$\\
Id	&	z	&$	(dm/dt)	$&$	(dM/dt)$&$	(\epsilon_{bol})	$&$	(\epsilon_j)	$&$	(\epsilon_j/\epsilon_{bol})	$&$	(\epsilon_{s/d})	$&$	(M_{dyn})	$\\
&&&$(M_{\odot}/yr)	$&&&&&$	(M_{\odot})	$\\
\hline
1	&$	0.0608	$&$	-1.16	$&$	-2.25	$&$	-1.54	$&$	-1.35	$&$	0.19	$&$	0.82	$&$	7.56	$\\
2	&$	0.1117	$&$	-0.68	$&$	-1.05	$&$	-0.91	$&$	-1.16	$&$	-0.25	$&$	0.53	$&$	8.28	$\\
3	&$	0.1628	$&$	-1.05	$&$	-1.86	$&$	-1.39	$&$	-1.24	$&$	0.15	$&$	0.82	$&$	7.84	$\\
4	&$	0.1316	$&$	-1.52	$&$	-1.65	$&$	-2.02	$&$	-1.60	$&$	0.42	$&$	0.93	$&$	8.52	$\\
\hline
\end{tabular}
\end{minipage}
\end{table*}

\begin{table*}
\begin{minipage}{165mm}
\contcaption{Mass Accretion Rates and Efficiency Factors for the NB16 LINERs.}   
\label{tab:6A}        
\begin{tabular}{lllllllll}   
\hline\hline                    
(1)&(2)&(3)&(4)&(5)&(6)&(7)&(8)&(9)\\
NB16&&$\rm{Log}$	&$\rm{Log}$&$\rm{Log}$&$\rm{Log}$&$\rm{Log}$&$	\rm{Log}$&$	\rm{Log}$\\
Id	&	z	&$	(dm/dt)	$&$	(dM/dt)$&$	(\epsilon_{bol})	$&$	(\epsilon_j)	$&$	(\epsilon_j/\epsilon_{bol})	$&$	(\epsilon_{s/d})	$&$	(M_{dyn})	$\\
&&&$(M_{\odot}/yr)	$&&&&&$	(M_{\odot})	$\\
\hline
60	&$	0.1273	$&$	-1.85	$&$	-2.01	$&$	-2.45	$&$	-0.21	$&$	2.24	$&$	2.64	$&$	8.49	$\\
61	&$	0.0959	$&$	-2.09	$&$	-2.28	$&$	-2.77	$&$	-0.31	$&$	2.46	$&$	2.78	$&$	8.46	$\\
62	&$	0.1077	$&$	-2.17	$&$	-2.26	$&$	-2.87	$&$	-1.12	$&$	1.75	$&$	2.04	$&$	8.56	$\\
63	&$	0.03	$&$	-1.78	$&$	-2.85	$&$	-2.37	$&$	-1.24	$&$	1.13	$&$	1.55	$&$	7.58	$\\
\hline
\end{tabular}
\end{minipage}
\end{table*}

\begin{table*}
\begin{minipage}{165mm}
\contcaption{Mass Accretion Rates and Efficiency Factors for the NB16 LINERs.}   
\label{tab:6A}        
\begin{tabular}{lllllllll}   
\hline\hline                    
(1)&(2)&(3)&(4)&(5)&(6)&(7)&(8)&(9)\\
NB16&&$\rm{Log}$	&$\rm{Log}$&$\rm{Log}$&$\rm{Log}$&$\rm{Log}$&$	\rm{Log}$&$	\rm{Log}$\\
Id	&	z	&$	(dm/dt)	$&$	(dM/dt)$&$	(\epsilon_{bol})	$&$	(\epsilon_j)	$&$	(\epsilon_j/\epsilon_{bol})	$&$	(\epsilon_{s/d})	$&$	(M_{dyn})	$\\
&&&$(M_{\odot}/yr)	$&&&&&$	(M_{\odot})	$\\
\hline
121	&$	0.0832	$&$	-1.50	$&$	-2.17	$&$	-1.99	$&$	-0.81	$&$	1.18	$&$	1.70	$&$	7.98	$\\
122	&$	0.0244	$&$	-1.96	$&$	-2.60	$&$	-2.60	$&$	-1.78	$&$	0.82	$&$	1.18	$&$	8.01	$\\
123	&$	0.0402	$&$	-1.77	$&$	-2.91	$&$	-2.34	$&$	-1.14	$&$	1.20	$&$	1.63	$&$	7.51	$\\
124	&$	0.1439	$&$	-1.50	$&$	-1.73	$&$	-1.99	$&$	-0.67	$&$	1.32	$&$	1.83	$&$	8.42	$\\
\hline
\end{tabular}
\end{minipage}
\end{table*}

\begin{table*}
\begin{minipage}{165mm}
\contcaption{Mass Accretion Rates and Efficiency Factors for the NB16 LINERs.}   
\label{tab:6A}        
\begin{tabular}{lllllllll}   
\hline\hline                    
(1)&(2)&(3)&(4)&(5)&(6)&(7)&(8)&(9)\\
NB16&&$\rm{Log}$	&$\rm{Log}$&$\rm{Log}$&$\rm{Log}$&$\rm{Log}$&$	\rm{Log}$&$	\rm{Log}$\\
Id	&	z	&$	(dm/dt)	$&$	(dM/dt)$&$	(\epsilon_{bol})	$&$	(\epsilon_j)	$&$	(\epsilon_j/\epsilon_{bol})	$&$	(\epsilon_{s/d})	$&$	(M_{dyn})	$\\
&&&$(M_{\odot}/yr)	$&&&&&$	(M_{\odot})	$\\
\hline
181	&$	0.1096	$&$	-2.03	$&$	-1.97	$&$	-2.70	$&$	-0.05	$&$	2.65	$&$	2.99	$&$	8.71	$\\
182	&$	0.1082	$&$	-1.96	$&$	-2.07	$&$	-2.60	$&$	-1.31	$&$	1.29	$&$	1.65	$&$	8.54	$\\
183	&$	0.0751	$&$	-2.25	$&$	-2.19	$&$	-2.98	$&$	-1.41	$&$	1.57	$&$	1.84	$&$	8.71	$\\
184	&$	0.1687	$&$	-1.34	$&$	-1.78	$&$	-1.78	$&$	-0.89	$&$	0.89	$&$	1.45	$&$	8.21	$\\
\hline
\end{tabular}
\end{minipage}
\end{table*}

\begin{table*}
\begin{minipage}{165mm}
\contcaption{Mass Accretion Rates and Efficiency Factors for the NB16 LINERs.}   
\label{tab:6A}        
\begin{tabular}{lllllllll}   
\hline\hline                    
(1)&(2)&(3)&(4)&(5)&(6)&(7)&(8)&(9)\\
NB16&&$\rm{Log}$	&$\rm{Log}$&$\rm{Log}$&$\rm{Log}$&$\rm{Log}$&$	\rm{Log}$&$	\rm{Log}$\\
Id	&	z	&$	(dm/dt)	$&$	(dM/dt)$&$	(\epsilon_{bol})	$&$	(\epsilon_j)	$&$	(\epsilon_j/\epsilon_{bol})	$&$	(\epsilon_{s/d})	$&$	(M_{dyn})	$\\
&&&$(M_{\odot}/yr)	$&&&&&$	(M_{\odot})	$\\
\hline
241	&$	0.1872	$&$	-1.25	$&$	-1.45	$&$	-1.66	$&$	-1.55	$&$	0.11	$&$	0.70	$&$	8.45	$\\
242	&$	0.1644	$&$	-1.50	$&$	-2.00	$&$	-1.99	$&$	-1.09	$&$	0.90	$&$	1.41	$&$	8.15	$\\
243	&$	0.1318	$&$	-0.84	$&$	-1.47	$&$	-1.11	$&$	-1.64	$&$	-0.53	$&$	0.20	$&$	8.02	$\\
244	&$	0.0432	$&$	-2.08	$&$	-2.26	$&$	-2.77	$&$	-0.59	$&$	2.18	$&$	2.50	$&$	8.47	$\\
\hline
\end{tabular}
\end{minipage}
\end{table*}

\begin{table*}
\begin{minipage}{165mm}
\contcaption{Mass Accretion Rates and Efficiency Factors for the NB16 LINERs.}   
\label{tab:6A}        
\begin{tabular}{lllllllll}   
\hline\hline                    
(1)&(2)&(3)&(4)&(5)&(6)&(7)&(8)&(9)\\
NB16&&$\rm{Log}$	&$\rm{Log}$&$\rm{Log}$&$\rm{Log}$&$\rm{Log}$&$	\rm{Log}$&$	\rm{Log}$\\
Id	&	z	&$	(dm/dt)	$&$	(dM/dt)$&$	(\epsilon_{bol})	$&$	(\epsilon_j)	$&$	(\epsilon_j/\epsilon_{bol})	$&$	(\epsilon_{s/d})	$&$	(M_{dyn})	$\\
&&&$(M_{\odot}/yr)	$&&&&&$	(M_{\odot})	$\\
\hline
302	&$	0.1701	$&$	-2.05	$&$	-2.35	$&$	-2.72	$&$	-0.72	$&$	2.00	$&$	2.34	$&$	8.35	$\\
303	&$	0.0939	$&$	-0.91	$&$	-1.73	$&$	-1.21	$&$	-1.75	$&$	-0.54	$&$	0.17	$&$	7.83	$\\
304	&$	0.0649	$&$	-1.93	$&$	-2.18	$&$	-2.56	$&$	-1.55	$&$	1.01	$&$	1.38	$&$	8.40	$\\
305	&$	0.0134	$&$	-1.88	$&$	-3.82	$&$	-2.49	$&$	-0.95	$&$	1.54	$&$	1.93	$&$	6.71	$\\
\hline
\end{tabular}
\end{minipage}
\end{table*}

\begin{table*}
\begin{minipage}{165mm}
\contcaption{Mass Accretion Rates and Efficiency Factors for the NB16 LINERs.}   
\label{tab:6A}        
\begin{tabular}{lllllllll}   
\hline\hline                    
(1)&(2)&(3)&(4)&(5)&(6)&(7)&(8)&(9)\\
NB16&&$\rm{Log}$	&$\rm{Log}$&$\rm{Log}$&$\rm{Log}$&$\rm{Log}$&$	\rm{Log}$&$	\rm{Log}$\\
Id	&	z	&$	(dm/dt)	$&$	(dM/dt)$&$	(\epsilon_{bol})	$&$	(\epsilon_j)	$&$	(\epsilon_j/\epsilon_{bol})	$&$	(\epsilon_{s/d})	$&$	(M_{dyn})	$\\
&&&$(M_{\odot}/yr)	$&&&&&$	(M_{\odot})	$\\
\hline
362	&$	0.0224	$&$	-2.42	$&$	-3.01	$&$	-3.20	$&$	-1.43	$&$	1.77	$&$	1.99	$&$	8.06	$\\
363	&$	0.0284	$&$	-2.37	$&$	-2.62	$&$	-3.13	$&$	-1.65	$&$	1.48	$&$	1.72	$&$	8.39	$\\
364	&$	0.0199	$&$	-2.45	$&$	-3.36	$&$	-3.26	$&$	-1.16	$&$	2.10	$&$	2.30	$&$	7.74	$\\
365	&$	0.0255	$&$	-2.29	$&$	-3.06	$&$	-3.04	$&$	-1.29	$&$	1.75	$&$	2.01	$&$	7.88	$\\
\hline
\end{tabular}
\end{minipage}
\end{table*}

\begin{table*}
\begin{minipage}{165mm}
\contcaption{Mass Accretion Rates and Efficiency Factors for the NB16 LINERs.}   
\label{tab:6A}        
\begin{tabular}{lllllllll}   
\hline\hline                    
(1)&(2)&(3)&(4)&(5)&(6)&(7)&(8)&(9)\\
NB16&&$\rm{Log}$	&$\rm{Log}$&$\rm{Log}$&$\rm{Log}$&$\rm{Log}$&$	\rm{Log}$&$	\rm{Log}$\\
Id	&	z	&$	(dm/dt)	$&$	(dM/dt)$&$	(\epsilon_{bol})	$&$	(\epsilon_j)	$&$	(\epsilon_j/\epsilon_{bol})	$&$	(\epsilon_{s/d})	$&$	(M_{dyn})	$\\
&&&$(M_{\odot}/yr)	$&&&&&$	(M_{\odot})	$\\
\hline
423	&$	0.1358	$&$	-1.56	$&$	-2.14	$&$	-2.06	$&$	-1.09	$&$	0.97	$&$	1.47	$&$	8.07	$\\
424	&$	0.0726	$&$	-1.83	$&$	-2.06	$&$	-2.44	$&$	-1.59	$&$	0.85	$&$	1.25	$&$	8.42	$\\
425	&$	0.0721	$&$	-1.37	$&$	-2.68	$&$	-1.83	$&$	-0.98	$&$	0.85	$&$	1.40	$&$	7.34	$\\
426	&$	0.1706	$&$	-2.07	$&$	-2.20	$&$	-2.75	$&$	-0.82	$&$	1.93	$&$	2.26	$&$	8.52	$\\
\hline
\end{tabular}
\end{minipage}
\end{table*}

\begin{table*}
\begin{minipage}{165mm}
\contcaption{Mass Accretion Rates and Efficiency Factors for the NB16 LINERs.}   
\label{tab:6A}        
\begin{tabular}{lllllllll}   
\hline\hline                    
(1)&(2)&(3)&(4)&(5)&(6)&(7)&(8)&(9)\\
NB16&&$\rm{Log}$	&$\rm{Log}$&$\rm{Log}$&$\rm{Log}$&$\rm{Log}$&$	\rm{Log}$&$	\rm{Log}$\\
Id	&	z	&$	(dm/dt)	$&$	(dM/dt)$&$	(\epsilon_{bol})	$&$	(\epsilon_j)	$&$	(\epsilon_j/\epsilon_{bol})	$&$	(\epsilon_{s/d})	$&$	(M_{dyn})	$\\
&&&$(M_{\odot}/yr)	$&&&&&$	(M_{\odot})	$\\
\hline
483	&$	0.1807	$&$	-1.61	$&$	-1.83	$&$	-2.14	$&$	-1.16	$&$	0.98	$&$	1.46	$&$	8.43	$\\
484	&$	0.102	$&$	-1.67	$&$	-1.88	$&$	-2.22	$&$	-0.40	$&$	1.82	$&$	2.27	$&$	8.44	$\\
485	&$	0.0849	$&$	-2.01	$&$	-1.98	$&$	-2.66	$&$	-1.47	$&$	1.19	$&$	1.54	$&$	8.68	$\\
486	&$	0.1291	$&$	-1.19	$&$	-2.07	$&$	-1.57	$&$	-1.19	$&$	0.38	$&$	1.00	$&$	7.77	$\\
\hline
\end{tabular}
\end{minipage}
\end{table*}

\begin{table*}
\begin{minipage}{165mm}
\contcaption{Mass Accretion Rates and Efficiency Factors for the NB16 LINERs.}   
\label{tab:6A}        
\begin{tabular}{lllllllll}   
\hline\hline                    
(1)&(2)&(3)&(4)&(5)&(6)&(7)&(8)&(9)\\
NB16&&$\rm{Log}$	&$\rm{Log}$&$\rm{Log}$&$\rm{Log}$&$\rm{Log}$&$	\rm{Log}$&$	\rm{Log}$\\
Id	&	z	&$	(dm/dt)	$&$	(dM/dt)$&$	(\epsilon_{bol})	$&$	(\epsilon_j)	$&$	(\epsilon_j/\epsilon_{bol})	$&$	(\epsilon_{s/d})	$&$	(M_{dyn})	$\\
&&&$(M_{\odot}/yr)	$&&&&&$	(M_{\odot})	$\\
\hline
545	&$	0.0421	$&$	-2.32	$&$	-2.33	$&$	-3.07	$&$	-1.69	$&$	1.38	$&$	1.63	$&$	8.64	$\\
546	&$	0.0809	$&$	-1.42	$&$	-2.45	$&$	-1.87	$&$	-1.12	$&$	0.75	$&$	1.29	$&$	7.61	$\\
547	&$	0.1115	$&$	-1.51	$&$	-1.92	$&$	-2.01	$&$	-1.36	$&$	0.65	$&$	1.16	$&$	8.24	$\\
548	&$	0.0635	$&$	-1.58	$&$	-2.38	$&$	-2.08	$&$	-1.26	$&$	0.82	$&$	1.32	$&$	7.84	$\\
\hline					
\hline
\end{tabular}
\end{minipage}
\end{table*}
\clearpage




\bsp	
\label{lastpage}
\end{document}